\input mnrass.sty
\pageoffset{-2.5pc}{0pc}
 

\Autonumber  


\pagerange{000--000}
\pubyear{1996}
\volume{000}

\begintopmatter  

\title{A new analysis of the Red Giant Branch \lq{Tip}\rq\ distance scale and the
value of the Hubble constant }

\author{ Maurizio Salaris$^{1}$ \& Santi Cassisi$^{1,2,3}$}

\affiliation{$^1$Max-Planck-Institut f\"ur Astrophysik, D-85740,
Garching, Germany - E-Mail: maurizio@MPA-Garching.mpg.de} 
\affiliation{$^2$Universit\'a degli studi de L'Aquila, Dipartimento di Fisica,
Via Vetoio, I-67100, L'Aquila, Italy}
\affiliation{$^3$Osservatorio Astronomico di Collurania, Via M. Maggini,
 I-64100, Teramo, Italy - E-Mail: cassisi@astrte.te.astro.it}

\shortauthor{M.Salaris \& S.Cassisi}
\shorttitle{Hubble constant from the TRGB distances}


\abstract

\tx
The theoretical evaluations of the Red Giant Branch Tip 
(TRGB) luminosity presented in Salaris \& Cassisi (1997)
are extended to higher metallicities, and compared with
analogous independent results recently published.
The present sets of stellar models agree
quite well in the determination of the TRGB brightness.

Relations between 
TRGB bolometric and I (Cousins) magnitude, and 
Zero Age Horizontal Branch V magnitude 
with respect to the metallicity are provided by adopting 
empirical, semiempirical and theoretical evaluations of bolometric corrections,
after a careful calibration of the zero point of the bolometric
correction scales.

The comparison between our ZAHB and TRGB distance
scales for galactic globular clusters presented in Paper I is now
supplemented with a comparison with the HIPPARCOS distance scale
set by local subdwarfs with accurate parallax determinations. 
The overall agreement between ZAHB and HIPPARCOS distances is quite good.
The TRGB distances for globular clusters are compatible with the ZAHB 
distances in the limit of the small sample of red giants observed.

The ZAHB and TRGB distances to resolved galaxies are in good agreement,
whereas the comparison
between TRGB and Cepheid distances, computed by
using the calibration suggested by Madore \& Freedman (1991), reveals a systematic
discrepancy of the order of 0.12 mag. The TRGB distances 
are systematically longer in comparison with the Cepheid ones.
This result supports the case for a revision of the 
zero point of the Cepheid distance scale, as already suggested by
other authors on the basis of HIPPARCOS parallaxes.
We do not find any clear correlation of the difference between TRGB and
Cepheid distances with metal content.

The application of our TRGB distance scale to NGC3379, provides a distance
to the Leo I group that is about 8\% higher than the one obtained by
Sakai et al. (1997a) adopting the TRGB-metallicity calibration by Lee,
Freedman \& Madore (1993). Our distance to the Leo I group, coupled 
with recent independent determinations of the distance Coma cluster-Leo I, 
obtained differentially by means of secondary distance indicators, 
provides a determination
of $H_{0}$ at the Coma cluster in the range: 
$H_{0}$= 60$\pm$11 Km $s^{-1} Mpc^{-1}$.

For choices of $\Omega$ in agreement with the observations
(0.3$\leq$$\Omega$$\leq$1) and cosmological constant equal to zero, 
our derived $H_{0}$ value is compatible with the most recent
determinations of the galactic globular clusters ages, thus removing
the long-standing conflict between the Hubble age and the age of the
oldest stars in the Galaxy.


\maketitle  


\section{Introduction}

\tx 
For determining the value of the Hubble constant ($H_{0}$) it is
necessary to measure with high accuracy distances to galaxies sufficiently
distant so that local departures from the Hubble flow are negligible.
This means that one needs to extend the distance scale to distances of
the order of 100 Mpc or more; 
to this aim different secondary distance indicators have been devised
in order to evaluate the relative distances between closer galaxies
and  more distant ones. However, the determination of $H_{0}$ 
is strongly dependent on the first step of the cosmological distance
ladder, that is constituted by the absolute distances of
close galaxies as determined by means of the primary
distance indicators. 

The Cepheid period-luminosity (P-L) relation is the
basis for the calibration of the extragalactic distance scale;
however, Cepheids observations are restricted only to
Population I systems and to late-type galaxies.
An excellent alternative primary distance indicator is the 
TRGB; its use results particularly
attractive since it is applicable to all morphological types of
galaxies as long as an old stellar population is present, and it has
been recently successfully applied by many authors for
estimating the distances to several nearby galaxies
(see e.g. Lee, Freedman \& Madore 1993, hereinafter LFM93, 
Sakai, Madore \& Freedman 1996; Soria et al. 1996; Elson 1996; Sakai
et al. 1997b).

The TRGB marks the helium ignition in the degenerate He core of
low-mass stars, and its luminosity depends on the He core mass,
which is remarkably constant for ages larger than 2-3 Gyr (see e.g.
Salaris \& Cassisi 1997, hereinafter Paper I), the exact value
depending on the metallicity.
Moreover, the I (Cousins) magnitude of the TRGB is only weakly sensitive
to the metallicity of the stellar population
and it is therefore obvious to use, as suggested by LFM93, the
observed I magnitude of the TRGB stars as a distance indicator.
LFM93 have provided a semiempirical calibration of this method 
in a large range of metallicity, and 
in a subsequent paper Madore \& Freedman (1995) undertook a series of numerical
simulations and concluded that the TRGB  method can be successfully used to
determine distances accurate to 0.2 mag for galaxies out to 3 Mpc using
ground based telescopes, and out to 12-13 Mpc using the Hubble Space Telescope.

Very recently Sakai et al. (1997a, hereinafter SA97) have determined,
by means of the TRGB method,
the distance to NGC3379 (an E1 galaxy at the center of the Leo I group)
and, then, taking advantage of the relative distance determination (obtained by means of 
secondary indicators) between the Leo I group and the Coma cluster,
they determined the absolute distance to the Coma cluster and 
$H_{0}=68\pm13 km s^{-1} Mpc^{-1}$.
The TRGB distance to NGC3379, crucial for the derived value of $H_{0}$,
has been obtained by adopting the LFM93 calibration of $M_{I}^{tip}$ 
as a function of [M/H] (as usual, we consider
[M/H]=log$(M/H)_{star}$-log$(M/H)_{\odot}$,
where M and H are, respectively, the global heavy element abundance and
the hydrogen abundance).
The LFM93 TRGB calibration is taken from Da Costa \& Armandroff (1990, 
hereinafter DA90). DA90 provides an empirical determination of the
bolometric correction for the I band, and a relation between the TRGB
bolometric magnitude and [M/H]; more in detail, DA90 took the
slope of the relation between 
TRGB brightness and the metallicity from the Sweigart \&
Gross (1978) models, while the zero point was set by
empirical TRGB $M_{Bol}$ values (from Frogel et al.1983), 
adopting the HB distance scale by Lee et al. (1990).
In addition to the fact that this calibration is partially based on
stellar models computed by adopting old physical inputs,
there is an important point that has to be taken
properly into account when using this calibration.
Frogel et al. (1983) determined the TRGB $M_{Bol}$ for a sample of
globular clusters (GC)
with different metallicities, observing only a small
sample of stars for each cluster.
As discussed in Paper I, when the sample of stars near the TRGB is
small, the TRGB brightness is on average underestimated; this causes therefore a zero
point too faint in the calibration of the TRGB magnitude with respect
to the metallicity.

In Paper I we have presented updated theoretical RGB stellar models
covering the metallicity range $-2.35\le[M/H]\le-0.57$, and we used
the computed TRGB luminosities for determining distances to a sample
of 11 nearby galaxies.
In this paper we will use the theoretical models presented in Paper I,
now extended to higher metallicities,
that take into account updated determinations of the physical 
inputs used in evolutionary computations.
By adopting this updated set of stellar models we 
firstly evaluate the distance modulus of NGC3379 on the basis of the TRGB,
(that will result higher than the value obtained by SA97)
and then the distance to the Coma cluster and the value of the Hubble constant.

In section 2 the results from the theoretical RGB models presented in Paper I are
briefly summarized and extended to a larger
metallicity range, together with a discussion about
the sets of color-transformations and bolometric corrections
adopted in the present work, that reproduces better observational
constraints. In section 3 we will compare our TRGB distance scale 
with other distance indicators in the light of the new sets of 
transformations adopted.
In section 4 the distance to Leo I is determined by means of the TRGB,
and by adopting a relative distance between Leo I and Coma as derived
from secondary distance indicators, the value of $H_{0}$ is finally obtained.
A summary and a discussion about the implication of our $H_{0}$
determination for the age of the Universe are presented in section 5.

\section{\bf The TRGB distance scale}
\subsection{\bf Theoretical stellar models}
\tx
The models used in this paper have been already presented
in Paper I. To summarize, we have determined the TRGB luminosities 
for stellar populations with age t=15 Gyr (but, as discussed
before and in Paper I, the precise value of t does not influence
the TRGB luminosities for ages larger than a few Gyr), metallicity
$-2.35\le[M/H]\le-0.57$ and Y=0.23, by computing evolutionary tracks of
low mass stars without chemical elements diffusion.
As far as it concerns the physical inputs adopted in computing the stellar models,
the interested reader is referred to Paper I.

To extend the TRGB luminosity calibration to higher metallicities, we
have now computed also evolutionary tracks for Z=0.01 and
Y=0.255 ([M/H]=-0.28), adopting the same input physics as in Paper I. 

Assuming for the Sun $M_{Bol,\odot}=4.75$ mag, in Paper I we gave a relation
between $M^{tip}_{Bol}$ and [M/H] that covered all the metallicity range
$-2.35\le[M/H]\le-0.57$. We have now verified, by computing stellar
models for the appropriate metallicity, that the same relation
reproduces the computed value of $M^{tip}_{Bol}$ at [M/H]=$-0.28$ within 0.02 mag;
therefore we can safely use it for a larger metallicity range: 

$${M^{tip}_{Bol}}=-3.949 - 0.178\cdot[M/H] + 0.008\cdot{[M/H]}^2 
\,\,\,\,\,\,\,\,\,\,(1)$$

\noindent
that covers the range $-2.35\le[M/H]\le-0.28$. 

It takes also
automatically into account the enhancement of the $\alpha$ elements
observed in galactic field halo and GC stars
(see, e.g. the review by Wheeler, Sneden \& Truran 1989) 
when considering the global metallicity [M/H].
In fact, as already demonstrated by Salaris, Chieffi \& Straniero
(1993) and verified by means of the models by Salaris \& Weiss (1997, 1998),
the TRGB bolometric magnitudes and ZAHB luminosities in the
[M/H] range spanned by the models presented in this paper 
as derived from $\alpha$-enhanced theoretical models, 
are well reproduced by scaled solar ones with the same global metallicity.
For fixed values of [$\alpha$/Fe]$\ge$0 and [Fe/H], the global metallicity [M/H]
is given by (see Salaris et al. 1993):

$$\rm [M/H]\approx[Fe/H] + log(0.638\cdot$f$+0.362)\,\,\,\,\,\,\,\,\,\,\,\,
\,\,\,\,\,\,\,\,\,\,\,\,\,\,\,\,\,\,\,\,\,\,\,\,\,\,\,\,\,(2)$$

\noindent
where log($f$)=[$\alpha$/Fe].

Equation 1 depends on the adopted initial Helium content (Y) since a variation of Y 
at fixed metallicity changes the TRGB luminosity because the change of the He core 
mass at the He flash. Regardless of the adopted
$\Delta{Y}/\Delta{Z}$ law and for reasonable choices about it, i.e. 
$1<\Delta{Y}/\Delta{Z}<5$ (see Peimbert 1993; Carigi et al. 1995), 
the He abundance that one has to adopt at $Z\le0.006$ is not substantially 
different from the cosmological value $Y\approx$0.23; the maximum variation is of about +0.03 
at Z=0.006 (for $\Delta{Y}/\Delta{Z}$=5). 
At [M/H]=$-0.28$ we have adopted Y=0.255 assuming $\Delta{Y}/\Delta{Z}$=2.5.
It is possible to take into account different Y abundances around the
values adopted in the present work by considering that on average 
${\partial{M_{Bol}^{tip}}}\over{\partial{Y}}$ is $\approx1.0$ in the
metallicity range covered by Equation 1.

In Figure 1, our prescription for the bolometric magnitude of the TRGB
as a function of [M/H] is displayed. For the aim of comparison, similar
relations as derived from the recent evolutionary models by
Cassisi et al. (1997 - their 'step8', with and without He and heavy
elements diffusion), Caloi 
et al. (1997 - no atomic diffusion) and Straniero et al. (1997 - no
atomic diffusion) are also plotted. 
These evolutionary models have been computed using slightly different
input physics with respect to our models (the reader is referred to
the quoted papers for more details on this subject), and in the case of the Caloi
et al. (1997) results, a completely independent evolutionary code was adopted.
In the same figure the relation 
provided by DA90 is shown after correcting for the slightly different $M_{Bol,\odot}$
adopted by the quoted authors.
The following points are worth noticing:
\smallskip
\noindent
i) the agreement between the different recent evolutionary results is
quite good. All the theoretical relations lie within $\approx \pm$0.05
mag with respect to equation 1; this provides an estimation
of the internal accuracy of $M_{Bol}^{TRGB}$ values provided by the current updated
theoretical scenario;
\smallskip
\noindent 
ii) the change of the TRGB luminosity due to the inclusion of
atomic diffusion - adopting the same physical
inputs as in standard models - is quite negligible (see the results
corresponding to the Cassisi et al. 1997 models);
\smallskip
\noindent
iii) the slope of the relation $M_{Bol}^{TRGB} - [M/H]$, provided
by the most recent stellar models is similar to the slope suggested by DA90;
\smallskip
\noindent
iv) there exists a difference of about 0.15 mag in the zero point between our
relation and the relation provided by DA90.
\smallskip
\noindent

\figure{1}{S}{80mm}{\bf Figure 1. \rm 
Comparison between updated predictions of stellar evolution theory, concerning the
behavior of the bolometric magnitude of the TRGB as a function of metallicity. The 
calibration provided by DA90 is also plotted. In all cases, a bolometric magnitude
for the Sun equal to 4.75 mag, has been adopted.}

\subsection{\bf The Bolometric Correction scale}

\tx
In order to derive the distance modulus through the TRGB method
LMF93 suggested an iterative procedure from observations in the 
VI Johnson-Cousins bands. Such procedure can be summarized as follows
(see LFM93 for more details):
\medskip
\noindent
i) fixing preliminarily the distance modulus;
\smallskip
\noindent
ii) with the fixed distance modulus determining the metallicity by
measuring the dereddened $(V-I)$ color at $M_{I}=-3.5$ mag
($(V-I)_{0,-3.5}$) and using a
relation between this color and the metallicity of the parent stellar population;
\smallskip
\noindent
iii) obtaining the distance modulus from the observed I magnitude of
the TRGB (corrected for the interstellar extinction) by adopting 
relations for both the TRGB bolometric magnitude as a function of metallicity 
and the bolometric correction to the I magnitude ($BC_{I}$);
\smallskip
\noindent
iv) iterating the previous steps until convergency is obtained between
the distance modulus at step (i) and the one obtained after step (iii).
Since the weak dependence of $M_{I}^{tip}$ on the metallicity,
convergence is generally achieved after one iteration.
\medskip
\noindent

When the TRGB bolometric magnitude is known from stellar models, the last 
ingredient necessary for the application of the TRGB method is
a relation providing the bolometric correction in the I (Cousins) band.
Following LFM93, an empirical $BC_{I}-(V-I)_{0}$ relation for RGB stars has been
taken from DA90. In that paper the authors give:

$$ BC_I= 0.881- 0.243\cdot(V-I)_0 \,\,\,\,\,\,\,\,\,\,\,\,\,\,\,\,\,\,
\,\,\,\,\,\,\,\,\,\,\,\,\,\,\,\,\,\,\,\,\,\,\,\,\,\,\,\,\,\,\,\,\,(3)$$

\noindent
independent of the metallicity.
The bolometric corrections provided by this relation are on a scale where
$BC_{V,\odot}$=-0.07. Since $M_{V,\odot}$=4.82$\pm$0.02 mag (Hayes 1985), this implies
the adoption of $M_{Bol,\odot}$=4.75 mag (as in Equation 1). We assume the quoted
error on the value of $M_{V,\odot}$ as an estimate of the error on the zero point of
the bolometric correction scale.

At this point it is straightforward to derive the distance modulus of a
galaxy according to the relation:

$$(m-M)_{I}=I_{TRGB}+BC_{I}-M^{tip}_{Bol} \,\,\,\,\,\,\,\,\,\,\,\,\,\,\,\,\,\,\,\,
\,\,\,\,\,\,\,\,\,\,\,\,\,\,\,\,\,\,\,\,\,\,\,\,\,\,\,\,\,(4)$$

The empirical $BC_{I}$ provided by equation 3 were derived
comparing the I magnitudes given in DA90 with the $M_{Bol}$ values
given by FPC83 for a sample of RGB stars in 8 GCs with different
metallicities. 
By examining Figure 14 in DA90, it appears clearly 
that in the range of $(V-I)_{0}$ values typical of the bulk of the stars considered 
by the authors ($(V-I)_{0}$ colors range from 1.0 to 1.6)
and of the TRGB stars in the sample of galaxies studied 
in the next sections (the $(V-I)_{0}^{TRGB}$ ranges approximately 
from 1.3 to 2.0) there is a dispersion of the order of 0.10 mag around the 
least-square fit given by Equation 3. Moreover, the relation for the
reddest stars is based only on a very small number of observational points.

In order to supply an independent determination of the $BC_{I}$ scale that can be
safely adopted for our RGB stellar models, we also used in Paper I the
theoretical $BC_{I}$ values derived from model atmospheres. 
More in detail, we used bolometric correction based on an updated version of 
the Kurucz's code ATLAS9 (Castelli 1996, private communication).
The $BC_{I}$ values were derived from the relation 
$BC_{I}=BC_{V}+(V-I)$, where $BC_{V}$ and $(V-I)$ are provided by the ATLAS9 code.
Obviously the zero point of the $BC_{V}$ scale sets also the 
zero point of the $BC_{I}$ scale. The Kurucz $BC_{V}$ are normalized in
such a way that the maximum value of $BC_{V}$ is zero, and all the
other values are negative. 
With this choice the value of $BC_{V,\odot}$ is
$\approx$-0.19, and therefore we should have used $M_{Bol,\odot}=4.63$ mag
for reproducing the observed $M_{V,\odot}$ value. However, we were
dealing with stars with a metallicity lower than the Sun, and
therefore we decided to adopt Vega ([M/H]=$-0.5$ according to Castelli \& Kurucz (1994)
for setting the zero point of our $BC_{V}$ scale.
The $BC_{V}$ value for Vega
provided by the ATLAS9 transformations presented a good agreement with the empirical
value provided by Code et al. (1976), and since Code et al. (1976) empirical
$BC_{V}$ are set on a scale where $BC_{V,\odot}=-0.07$, we adopted in
Paper I the ATLAS9 transformations and $M_{Bol,\odot}=4.75$ mag.

However, a detailed comparison with the recent empirical 
$BC_{V}$ database provided by Alonso et al. (1995 - 
that is an extension and improvement of the work by Code et al. 1976) based on
observations
of many solar metallicity and metal poor Main Sequence stars, 
reveals that a value $M_{Bol,\odot}=4.62$ mag should be used for fitting the empirical
bolometric correction scale.
By assuming $M_{Bol,\odot}$=4.62 mag,
we had finely reproduced the results for a large sample of metal poor stars 
and also the observed value for $M_{V,\odot}$.

Bearing in mind this comparison (see also the discussion in De Santis 1996), 
in the present work we have decided to reanalize the results obtained in Paper I.
Once again, we have adopted the theoretical bolometric corrections and colors
obtained with the Kurucz's code ATLAS9, but now we have taken
advantage of a new grid of model atmospheres computed with an
updated version of the code (Castelli 1997, private communication; 
Castelli, Gratton \& Kurucz 1997a, 1997b, hereinafter K97).
We have verified that by adopting for $M_{Bol,\odot}$ the value 4.62 mag, with these
new model atmospheres we can simultaneously match both $M_{V,\odot}$ and the $M_{V}$
values given by the empirical $BC_{V}$ scale of Alonso et al. (1995).
By using our stellar models and the K97 $BC_{V}$ and colors, the following relation
has been obtained for the absolute I magnitude of the TRGB:

$$M_{I}^{tip}=-3.953 + 0.437\cdot[M/H] + 0.147\cdot[M/H]^2 \,\,\,\,\,\,\,\,\,\,\,\,\,(5)$$

\noindent
with a correlation coefficient r=0.99.
 
For estimating the uncertainty related to the use of theoretical bolometric corrections
and temperature conversion, being aware of the problems still existing with 
model atmospheres, we have also searched for semiempirical relations which satisfy 
observational constraints, indipendently from the DA90 $BC_{I}$.
For this reason we adopt here also the so called Yale transformations (Green 1988)
for obtaining the theoretical $BC_{I}$ values. These transformations
are an empirical UBVRI recalibration (independent of the DA90 $BC_{I}$ scale)
of Vandenberg \& Bell (1985) and Kurucz (1979) synthetic colors and
$BC_{V}$, taking into account various observational constraints. The
$BC_{V}$ values, based on a scale in which $BC_{V,\odot}=-0.07$,
are in satisfactory agreement (within less than 0.04-0.05 mag, when one takes into account
the difference in the assumed value of $BC_{V,\odot}$) with the more recent empirical
$BC_{V}$ by Alonso et al. (1995).
By adopting these transformations (together with $M_{Bol,\odot}=4.75$ mag)
and our theoretical models, we get the following relation:

$$M_{I}^{tip}=-4.156 + 0.157\cdot[M/H] + 0.070\cdot[M/H]^2 \,\,\,\,\,\,\,\,\,\,\,\,\,(6)$$

\noindent
with r=0.98.

In the next section we will check the consistency between the
empirical $BC_{I}$ given by DA90, the semiempirical ones from 
the Yale transformations and the theoretical $BC_I$ supplied by K97.

\section{\bf Comparison between TRGB, RR Lyrae and Cepheid distance scales}

\tx

Before using the TRGB method for determining the distance to the Leo I
group, in order to assess the reliability of the theoretical TRGB luminosities,
we will briefly compare in this section the TRGB distances with the distance scales 
set by RR Lyrae and Cepheids in GCs 
and nearby galaxies in which the stellar component has been resolved.

The major improvement in comparison with Paper I
is that now we also adopt the Yale and the K97 transformations, after a careful
calibration of the bolometric correction zero point as discussed in the
previous section. 
Moreover we will compare our RR Lyrae distances to GC
also with the distances obtained by means of the Main Sequence Fitting (MSF)
technique based on nearby subdwarfs for which accurate parallaxes have been recently 
measured by HIPPARCOS.

In the following we will separately discuss the cases for GCs and for resolved galaxies.

\subsection{\bf Globular Clusters}

\tx

In the case of galactic GCs it is possible to compare the distance scale fixed
by ZAHB models, with the one derived from Equation 1.
Here we have adopted the ZAHB models from Cassisi \& Salaris
(1997) and used in Paper I, but transformed into the observational plane
by using both the Yale and K97 transformations.
The relations between the ZAHB V magnitude (taken at
$\log{T_{eff}}=3.85$, that corresponds approximately to the average
temperature of the RR Lyrae instability strip)
and [M/H] are the following:

$$\rm M^{zahb}_{V, Yale}= 0.921 + 0.329\cdot[M/H] + 
0.045\cdot[M/H]^2\,\,\,\,(7)$$ 

$$\rm M^{zahb}_{V, K97}= 0.974 + 0.379\cdot[M/H] + 
0.062\cdot[M/H]^2\,\,\,\,(8)$$

\noindent
for $-2.35\le[M/H]\le-0.57$, with r=1.00 for both relations. 

The difference between these two ZAHB distance scales is quite negligible, being on average equal 
to 0.02-0.03 mag, the ZAHB luminosities obtained by using the Yale transformations  
being systematically brighter.
>From now on we will adopt Equation 7 as our reference ZAHB RR Lyrae distance
scale. 
  
The comparison between the TRGB and the ZAHB distance scales fixed by Equations 1 and 7 
has been performed, as in Paper I, by adopting the TRGB observational data by
Frogel et al. (1983 - hereinafter FPC83 - and reference therein), who 
provided absolute bolometric magnitudes (on a scale where $BC_{V,\odot}$=-0.07)
for many TRGB of galactic GCs. These magnitudes
have been obtained empirically by directly integrating the flux from the program stars
via the observed {\sl UBVJHK} photometry and adopting a RR Lyrae distance scale for 
the selected clusters. 

As far as it concerns the criteria adopted for selecting the clusters in our sample,
a detailed discussion can be found in Paper I.
Table 1 lists, for all clusters in our sample, the values of $[Fe/H]$ and $[{\alpha}/Fe]$ 
obtained by means of spectroscopic analysis (as collected by Salaris \& Cassisi 1996), 
the global metallicity [M/H] (according to relation 2), reddening, the distance modulus 
(reddening corrected) and $M_{Bol}^{tip}$ (obtained by modifing the values given by
FPC83 for taking into account the ZAHB distance scale given by our Equation 7, and
the reddenings and [M/H] values we adopt).

\table{1}{S}{\bf Table 1. \rm Selected data for the sample of galactic
globular clusters.} 
{\halign{%
\rm#\hfil& \hskip7pt\hfil\rm#\hfil &\hskip7pt\hfil\rm#\hfil &\hskip7pt\hfil\rm#\hfil &\hskip7pt\hfil\rm#\hfil
&\hskip7pt\hfil\rm#\hfil&\hskip7pt\hfil\rm\hfil#&\hskip7pt\hfil\rm\hfil#\cr
Cluster& [Fe/H] & $[\alpha/Fe]$ & [M/H] &E(B-V)&$(m-M)_{0}$ & $\rm {M}^{tip}_{Bol}$ \cr 
\noalign{\vskip 10pt}
M71    & -0.80 & 0.27 & -0.61 & 0.28 & 13.08 & -3.74 \cr 
NGC6352& -0.80 & 0.13 & -0.70 & 0.21 & 13.90 & -4.01 \cr
47 Tuc & -0.80 & 0.15 & -0.70 & 0.04 & 13.37 & -3.83 \cr   
NGC362 & -1.20 & 0.23 & -1.04 & 0.06 & 14.67 & -3.45 \cr
M5     & -1.40 & 0.30 & -1.19 & 0.03 & 14.47 & -3.37 \cr
M79    & -1.42 & 0.21 & -1.27 & 0.00 & 15.78 & -3.70 \cr
NGC6752& -1.50 & 0.31 & -1.28 & 0.04 & 13.16 & -3.60 \cr
M3     & -1.49 & 0.26 & -1.31 & 0.00 & 15.19 & -3.53 \cr
NGC6397& -1.88 & 0.25 & -1.70 & 0.18 & 11.99 & -3.36 \cr
M68    & -1.92 & 0.20 & -1.78 & 0.07 & 15.03 & -3.50 \cr
M15    & -2.30 & 0.30 & -2.09 & 0.10 & 15.19 & -3.53 \cr}}

Figure 2 shows the $M_{Bol}^{tip}$ values versus the global heavy elements
abundance for the selected clusters, and also the theoretical relation
for the TRGB luminosity (Equation 1).
The vertical error bar ($\pm0.1$ mag) for the observational points 
represents an average error on the distance modulus obtained from
relation 7 (see Cassisi \& Salaris 1997) while
the error on the spectroscopic determination of [M/H] is typically of the order of
0.15 dex (see Paper I).
\figure{2}{S}{100mm}{\bf Figure 2. \rm 
The absolute bolometric magnitude of the brightest observed red giant
as a function of the global
metallicity, for the sample of clusters selected from the FPC83 database. 
The solid line shows the theoretical expectation for the bolometric
magnitude of the TRGB. 
The dashed lines represent the same theoretical relation but
shifted in steps of 0.1 mag.}

Data in Figure 2 show quite clearly that the TRGB
observational points (with the exception of NGC6352)
are located at lower luminosities in comparison with the
theoretical relation, with an average difference of approximately 0.15-0.20 mag.
However, this is exactly what is expected on the basis of simple statistical
arguments (see the detailed discussion in Paper I), as soon as
the evolutionary times in the upper part of the RGB and the
number of stars observed in each cluster by FPC83 are taken into account.
This means that the theoretical TRGB and ZAHB distance scales in GCs
are in agreement within the statistical uncertainties due to the small sample of
red giant stars observed.

The metal-rich GC NGC6352 is the cluster out of the whole sample which
is characterized by a luminosity level higher than the theoretical value. 
When checking for it in the FPC83 paper
one notices that the star considered to be at the TRGB in this cluster 
could be a field star. If this is the case the second brightest star
in the FPC83 sample
is $\approx$0.3 mag fainter, and it would be located in Figure 2 below
the line corresponding to the theoretical TRGB values, as expected
from statistics.

As an independent check of the reliability of the distance moduli derived from 
Equation 7 and of the calibration of the evolutionary models,
we show in Figure 3, as an example, a fit to the I-(V-I) diagrams by
DA90 of the RGB in NGC6397 (lower panel) and in NGC6752 (upper panel) by
using our theoretical RGB models together with the Yale transformations, 
the $(m-M)_{0}$ and E(B-V) values
given in Table 1, and the extinction relations by Cardelli et al.
(1989). The agreement between theory and observations is quite
satisfactory. 

\figure{3}{S}{100mm}{\bf Figure 3. \rm 
Fit to the I-(V-I) diagrams of NGC6752 (upper panel) and NGC6397 (lower
panel) by DA90 with the theoretical RGB models presented in this paper. 
The cluster metallicities, distance moduli, and reddenings adopted in the fit are 
displayed.}

As a second check about our theoretical ZAHB distance scale for GCs we have
compared our results with GC distance moduli taken from the very recent literature, 
derived from the MSF technique based on subdwarfs with accurate HIPPARCOS parallaxes.
The sources of the GC MSF distances are Gratton
et al. (1997 - we have considered the
distances obtained by correcting for binary contamination to the
HIPPARCOS subdwarfs sample, as displayed in column 8 of their Table 3), Reid
(1997 - we have considered
the dereddened distance moduli displayed in column 7 of his Table 3,
derived assuming the reddening used in our paper,
and then applied the corresponding extinction contribution) and
Chaboyer et al. (1997).

In Figure 4 (panels a-c) we display the result of this comparison,
where the error bars on the MSF distances are taken from the
quoted papers.
Since the various authors adopt different 
procedures for the MSF, different sets of
subdwarfs, different corrections for the statistical bias affecting the subdwarfs
luminosities, and sometimes slightly different
assumptions about the [M/H] values for both the clusters and the subdwarfs, 
the differences between the distance moduli obtained for the GCs in common
among these three investigations give us a rough estimate of the intrinsic error 
of the MSF technique.
To make the comparison more meaningful, we have decided to adopt for each cluster
the same [Fe/H] values used by each of the quoted authors, and as
in Gratton et al. (1997),
[$\rm \alpha$/Fe]=0.30 (very similar to the average [$\rm \alpha$/Fe] values
from Table 1) for all the three cases.

From Figure 4 one can easily notice that on average there is a good agreement
between our ZAHB distance scale and the HIPPARCOS MSF distances.
For the most metal poor (and more distant) clusters displayed in the
figure, the Reid (1997) data seem to disagree systematically with our ZAHB
distances, but it is worth noting that the same M68 distance derived
by Gratton et al. (1997) nicely agrees with the ZAHB distance.
In particular, in the case of M5 and NGC6752 the distance moduli derived from the MSF 
by the three different groups are almost identical,
and the agreement with the ZAHB distance scale is almost perfect.

We can therefore finally conclude that for GCs the HIPPARCOS, TRGB and
ZAHB distance scales are in agreement one with each other within
the present errors.

With the distance scale set by Equation 7, it is now possible to
recalibrate Equation 7 of Paper I, which provides [M/H] as a 
function of $(V-I)_{0,-3.5}$, based on a subsample of the GCs listed in
table 1 for which DA90 provide V-(V-I) diagrams. 
The new relation is 

$$[M/H]=-39.270 + 64.687\cdot[(V-I)_{0,-3.5}]  {\hskip 2.5truecm}$$
$$\hskip 0.5truecm - 36.351\cdot[(V-I)_{0,-3.5}]^2+ 6.838\cdot[(V-I)_{0,-3.5}]^3  
\,\,\,\,\,(9)$$

\noindent
with r=0.99. As in Paper I, this relation has been obtained by imposing that the 
$(V-I)_{0,-3.5}$ values have to be monotonously increasing for increasing
metallicity.

\figure{4}{S}{120mm}{\bf Figure 4. \rm Comparison between different GC distance moduli, obtained by using
the MSF with HIPPARCOS subdwarfs and our ZAHB distance scale (Equation 7). In each panel, 
the ZAHB distance modulus has been computed by using the same [Fe/H] values
adopted by the corresponding MSF author and [$\alpha$/Fe]=0.30 (see
text for more details).} 

\figure{5}{S}{100mm}{\bf Figure 5. \rm Comparison between different TRGB distances
obtained by adopting the DA90 $BC_{I}$ with the results from LFM93}
\figure{6}{S}{100mm}{\bf Figure 6. \rm Comparison between different distances for the selected
sample of resolved galaxies, obtained using the TRGB together with the
DA90, Yale and K97 $BC_{I}$ values}
\figure{7}{S}{100mm}{\bf Figure 7. \rm Comparison between different distances for the selected
sample of resolved galaxies, obtained using the TRGB (by means of the DA90, Yale and
K97 $BC_{I}$)
and the RR Lyrae (equation 7) distance scales (filled circles). 
The error bars associated to each individual point are also displayed. 
The comparison
between the distance moduli obtained using the TRGB method and the RR Lyrae
distance scale but adopting for the RR Lyrae stellar population an average metallicity
equal to [M/H]=-1.5 (open circles) is also shown (see text).}
\table{2}{D}{\bf Table 2. \rm Selected parameters for a sample of resolved 
galaxies (see text).} 
{\halign{%
\rm#\hfil & \hskip3pt\hfil\rm#\hfil &\hskip3pt\hfil\rm#\hfil 
&\hskip3pt\hfil\rm#\hfil&
\hskip3pt\hfil\rm#\hfil &\hskip4pt\hfil\rm#\hfil & \hskip4pt\hfil\rm#\hfil & 
\hskip4pt\hfil\rm#\hfil & \hskip4pt\hfil\rm#\hfil & \hskip4pt\hfil\rm\hfil#\hfil\cr
Galaxy & $E(B-V)$ & $I_{TRGB}$ & $[M/H]$ & $(m-M)_{0,TRGB}^{DA90}$ & $(m-M)_{0,Ceph}$ & 
$(m-M)_{0,RR}$ & $(m-M)_{0,RR}^{-1.5}$ & $(m-M)_{0,TRGB}^{Yale}$  &
$(m-M)_{0,TRGB}^{K97}$ \cr 
\noalign{\vskip 10pt}
LMC     & 0.10  &14.60 &-1.0&18.60&18.50 &18.54 &      & 18.64& 18.64\cr
NGC6822 & 0.28  &20.05 &-1.7&23.61&23.62 &      &      & 23.68& 23.73\cr
NGC185  & 0.19  &20.30 &-1.0&24.12&      &24.06 &24.15 & 24.15& 24.15\cr
NGC147  & 0.17  &20.40 &-0.9&24.27&      &24.17 &24.28 & 24.29& 24.29\cr
IC1613  & 0.02  &20.25 &-1.2&24.43&24.42 &24.41 &24.46 & 24.45& 24.47\cr
M31     & 0.08  &20.55 &-0.9&24.56&24.44 &24.56 &24.67 & 24.62& 24.62\cr
M33     & 0.10  &20.95 &-2.0&24.82&24.63 &24.85 &24.78 & 24.92& 24.97\cr
WLM     & 0.02  &20.85 &-1.5&24.97&24.92 &      &      & 25.03& 25.08\cr
NGC205  & 0.035 &20.45 &-0.9&24.54&      &24.90 &25.01 & 24.62& 24.61\cr 
Sex A   & 0.03  &21.79 &-1.9&25.88&25.85 &      &      & 25.92& 25.97\cr
Sex B   & 0.015 &21.60 &-1.6&25.72&25.69 &      &      & 25.81& 25.85\cr
NGC3109 & 0.04  &21.55 &-1.5&25.61&25.60 &      &      & 25.69& 25.74\cr}}
\table{3}{S}{\bf Table 3. \rm Estimates of the individual errors associated to the
TRGB, Cepheid and RR Lyrae distances for the sample of galaxies in table 2 (see text).} 
{\halign{%
\rm#\hfil & \hskip3pt\hfil\rm#\hfil & \hskip3pt\hfil\rm#\hfil & \hskip3pt\hfil\rm#\hfil\cr
 Galaxy & $\Delta{(m-M)_{0,TRGB}}$  & $\Delta{(m-M)_{0,Ceph}}$ & $\Delta{(m-M)_{0,RR}}$   \cr 
\noalign{\vskip 10pt}
LMC    &  0.14  &           &    0.12 \cr
NGC6822  &  0.14  &     0.17  &         \cr
NGC185   &  0.32  &           &    0.20 \cr
NGC147   &  0.14  &           &    0.20 \cr
IC1613 &  0.22  &     0.13  &    0.20 \cr
M31    &  0.18  &     0.10  &    0.20 \cr
M33    &  0.18  &     0.09  &    0.20 \cr
WLM    &  0.14  &     0.11  &         \cr
NGC205   &  0.22  &           &    0.20 \cr
Sex A   &  0.14  &     0.15  &         \cr
Sex B   &  0.13  &     0.25  &         \cr
NGC3109  &  0.14  &     0.15  &         \cr}}

\subsection{\bf Resolved galaxies}

\tx

When considering resolved galaxies we can compare (as in Paper I, LFM93)
our TRGB distance moduli with RR Lyrae and Cepheid distances.
The observational database used in this comparison is the same one as in Paper I, 
with the additional data for Sextans B taken from Sakai et al. (1997b)
and new BVRI data for the Cepheids in NGC3109 from Musella et al. (1997). 
The Cepheid distance scale is the one set by the P-L relations
provided by Madore \& Freedman (1991),
with the zero point set by a LMC distance modulus of 18.50 mag and $\rm E(B-V)=0.10$.
The extinction law (Cardelli et al. 1989), adopted for correcting the apparent RR Lyrae and
TRGB distance moduli for the extinction, is the same adopted by
Madore \& Freedman (1991) and employed in all the Cepheid distance determinations
used in this comparison.

In Table 2 we report the distance modulus determinations as obtained with the 
three different methods. 
The various columns provide the following data: (1) the name of the object;
(2) the reddening; (3) the observed I magnitude of the TRGB;
(4) the mean RGB metallicity, as obtained by adopting Equation 9 and the
distance moduli in column 5;
(5) the true distance modulus estimated by using the TRGB method
and the DA90 $BC_{I}$; (6) the intrinsic Cepheid
distance; (7) the true distance obtained by using the mean RR Lyrae luminosity; 
(8) as in column (7) but for an average metallicity of the RR Lyrae population
[M/H]=-1.5 (see below); (9) the distance modulus obtained by applying
the TRGB method and
Equation 6 (the metallicities derived in this way are on average
$\approx$0.05-0.10 dex lower than the values in column 4); (10) the
distance modulus obtained applying the TRGB method and Equation 5
(also in this case the metallicities are on average
$\approx$0.05-0.10 dex lower than the values in column 4).

Estimates of the individual errors 
associated to the Cepheid, TRGB and RR Lyrae distances are given in Table 3. 
The errors on $(m-M)_{0,Cepheid}$ are taken from the corresponding papers
(in the case of NGC3109 we have used the new Cepheid observations
by Musella et al. (1997) and applied the method outlined in
Madore \& Freedman (1991) for deriving the distance and the related error)
without taking into account the contribution due to the uncertainty in
the adopted zero point of the P-L relation (the distance modulus of the LMC, set
at $(m-M)_{0}$=18.50) since we want to check if there
exists a discrepancy between the Cepheid distances set by this zero
point and the TRGB and RR Lyrae theoretical distance scales.

As for the errors on the TRGB distances, we have considered the error
associated to the detection of $I_{TRGB}$ as given in the original
papers, to this
it has been added statistically the contribution due to
an error on the metallicity of the parent stellar population
(as derived by means of the $\rm (V-I)$ RGB color) by $\pm$0.40 dex, an indetermination
on the initial He content Y by $\pm$0.03, an indetermination on the
theoretical calibration of the TRGB luminosity as given in section 2.1,
and an error on the zero point of each of the three different bolometric correction
scales adopted here as given in section 2.2. Moreover, it has been also
accounted for the contribution due to the uncertainty on the extinction: in
the case it is discussed in the original papers, we have adopted the value
given by the authors; if the value for the reddening
is taken from Burstein \& Heiles (1984), we have considered the 
correspondent error as given by the quoted authors, and then
translated it into an error on the extinction by adopting the reddening law
previously quoted.

In the case of RR Lyrae distances, the errors were derived from the 
observational errors on the mean brightness of the RR Lyrae sample
observed in each individual galaxy, and by adding statistically the contribution
due to the error related to the procedure followed for converting this
mean brightness of the RR Lyrae sample into the corresponding ZAHB
magnitude (see Paper I for details about the procedure).
We just notice that in this paper, when RR Lyrae observations have been performed
in the $g$ Thuan-Gunn band (as in the case of NGC185, NGC147, IC1613, NGC205),
the relation by Kent (1985) for transforming the $g$ magnitudes
into V magnitudes has been adopted.

It is also important to remember that
the [M/H] values used for obtaining the distance moduli given in column 7 of Table 2 
are derived from
RGB stars, and correspond to an average metallicity of this stellar population. In principle 
this metallicity could not correspond to the RR Lyrae average metal
content, especially for highest and lowest values of [M/H] displayed
in Table 2,  
due to the low probability that metal-poor and 
metal-rich RGB stars evolve during their He central burning phase
through the RR Lyrae instability strip
(as for example in the case of the metal poor GC M92 and the metal
rich one 47 Tuc).
For roughly estimating the uncertainty introduced by the unknown metallicity of the
RR Lyrae population, the distance moduli obtained assuming for the RR Lyrae stars 
an average metallicity equal to [M/H]=$-1.5$ - adopted as a reasonable estimate of the average
metallicity for the galactic GC RR Lyrae population - have been also reported 
(with the unique exception of the LMC; in this case the correct
metallicity for the considered RR Lyrae stars is taken into account,
see Paper I) in Table 2 (column 8).

\figure{8}{S}{100mm}{\bf Figure 8. \rm Comparison between different distances for the selected
sample of resolved galaxies, obtained by using the TRGB (by means of the
DA90, Yale and K97 $BC_{I}$) and the Cepheid distance scales. The
error bars associated to each individual point are also displayed. 
The long dashed line in each panel
corresponds to the weighted average difference between TRGB and Cepheid
distance moduli. Its associated error is also shown.}

Figure 5 shows the difference between the TRGB distance moduli
obtained by adopting the DA90 $BC_{I}$ and the corresponding results from
LFM93 (where the same $BC_{I}$ are used). It is clear from the
figure that there is a systematic offset
by on average 0.15 mag, our distances being larger.

We have then displayed in Figure 6 the difference between the TRGB distance
moduli obtained adopting respectively the DA90, Yale or K97 $BC_{I}$. 
The average difference $(m-M)_{TRGB,DA90}-(m-M)_{TRGB,Yale}$ is equal
to -0.06 mag, while the $(m-M)_{TRGB,DA90}-(m-M)_{TRGB,K97}$ is equal to -0.08mag. 

The difference between the TRGB distances - obtained by adopting alternatively 
the DA90, the Yale, or the K97 $BC_{I}$ - and the RR Lyrae distances is
shown in Figure 7. 
In all cases there is no statistically significant correlation
between $(m-M)_{TRGB}-(m-M)_{RRLyrae}$ and $(m-M)_{RRLyrae}$.
By using the DA90 $BC_{I}$ for the TRGB theoretical luminosity
and by neglecting the very discrepant point corresponding
to NGC205 - see also the discussion in LFM93 about this galaxy -,
if one considers for the RR Lyrae stars the same
metallicity as derived from the RGB stars, 
a weighted average difference $(m-M)_{TRGB}-(m-M)_{RRLyrae}$=0.03$\pm$0.10 mag has been
obtained. 
When considering (with the exception of the LMC) an average 
metallicity $\rm [M/H]=-1.5$, a weighted average difference 
$(m-M)_{TRGB}-(m-M)_{RRLyrae}$=-0.01$\pm$0.10 mag is derived.
 
In the case of the Yale or the K97 bolometric corrections, 
the same quantity $(m-M)_{TRGB}-(m-M)_{RRLyrae}$ 
ranges between 0.05$\pm$0.10 and 0.09$\pm$0.10 mag 
depending on the assumed RR Lyrae metallicity. 
One can therefore conclude that the RR Lyrae and TRGB distance scales
agree quite well. Their average difference is statistically
consistent with a value equal to zero when considering our sample of
resolved galaxies, spanning a metallicity range (for the RGB
stellar populations) by $\approx$1 dex.

It is worth noting the large difference which exists between our evaluation
(18.54 mag) and the LFM93 result (18.28 mag) for the LMC distance modulus
based on the RR Lyrae distance scale.
The origin of such discrepancy has to be related both to the difference 
($\approx0.1$ mag) between our RR Lyrae distance scale and the one adopted 
by LFM93, and to the different observational data for the LMC clusters,
we adopt in present work (see Paper I for a detailed discussion on this point).

\figure{9}{S}{100mm}{\bf Figure 9. \rm Average difference 
$(m-M)_{0,TRGB}-(m-M)_{0,Ceph}$ for the seven irregular galaxies
quoted in the text, as a function of the [O/H] abundance
in their HII regions (see text for more details).}

In Figure 8 (panels a-c)
we display the difference between the distance moduli obtained by adopting
the TRGB and the Cepheid distance scale. 
Also in this case there is no statistically significant correlation between
$(m-M)_{TRGB}-(m-M)_{Cepheid}$ and $(m-M)_{Cepheid}$. The points
corresponding to the individual galaxies are always located above the
line corresponding to a difference equal to zero.
The weighted average differences $(m-M)_{TRGB}-(m-M)_{Cepheid}$ obtained by adopting 
for the TRGB distance scale the Equation 1 and the DA90 $BC_{I}$, or Equation 6 or
Equation 5, are equal to 0.08$\pm$0.07 mag (very similar to the result 
obtained in Paper I with a smaller galaxy sample), 0.13$\pm$0.07 mag 
and 0.16$\pm$0.07 mag, respectively
(corresponding to the long dashed line in each panel of the Figure 8).
It is interesting to notice that in all of the three cases the systematic difference 
from zero is statistically significant.

The case of Sex B deserves a brief comment. According to Sakai et al. (1997b)
the Cepheid distance modulus is 25.69$\pm$0.25 mag (as displayed in
table 2) but, as discussed
in their paper, if one short period Cepheid in their sample
(possibly an overtone pulsator) is excluded from the analysis, it is possible
to find a solution with a reddening equal to zero and a distance
modulus equal to 25.82 mag. In this case, the difference
$(m-M)_{TRGB}-(m-M)_{Cepheid}$ for Sex B would be quite different.
We have therefore recomputed the average differences
$(m-M)_{TRGB}-(m-M)_{Cepheid}$ excluding Sex B or adopting the higher
value for its distance modulus. However in both cases the average differences
$(m-M)_{TRGB}-(m-M)_{Cepheid}$
for all the sample of galaxies are changed by not more than 0.01 mag.

If one considers the full range of values obtained by adopting the three
different sets of $BC_{I}$ as an estimate of the uncertainty in the
$BC_{I}$ scale, we derive an mean difference 
$(m-M)_{TRGB}-(m-M)_{Cepheid}$=0.12 mag.
It is interesting to note that this difference between TRGB and
Cepheid distances is in agreement
with very recent results obtained by adopting
HIPPARCOS parallaxes, which show that the LMC distance modulus (and therefore the zero
point of the Cepheid calibration) could be larger than 18.50.
LMC distance moduli equal to 18.60$\pm$0.07 mag, 18.65$\pm$0.07 mag,
18.60$\pm$0.20 mag,
18.70$\pm$0.10 mag, 18.56$\pm$0.08 mag are derived respectively by 
Gratton et al. (1997), Reid (1997), Whitelock et al. (1997), Feast \& Catchpole (1997), 
Oudumaijer et al. (1997), by adopting different calibrators and different techniques.

When considering these results, it is worth to bear in mind 
that the TRGB luminosities determined for 
the sample of galaxies displayed in Table 2 are based on observations of a very large
number of RGB stars, much larger than in the case of the FPC83
observations of GC RGB. Therefore,
according to the discussion presented in Paper I and following the 
results of the statistical analysis performed by Madore \& 
Freedman (1995),
in the case of the RGB star sample observed for each individual galaxy
we can compare directly the observed and predicted TRGB $I$ luminosities,
without any statistical uncertainty due to the small number of
observed stars, provided that the considered star sample contains \lq{real\rq\ RGB
stars belonging to the target galaxy.
However, one has always to be aware of the fact that in observations of these galaxies,
crowding and potential contaminants (as background galaxies, foreground stars, an AGB 
stellar population) have a systematic effect on the determination of $I_{TRGB}$, and that 
they have to be carefully treated for obtaining a reliable TRGB determination.
According to Madore \& Freedman (1995) the influence of background/foreground 
contamination can be strongly reduced when working as far out in the halo of the
galaxies as possible. They give also a set of criteria and a
method for deriving consistently TRGB luminosities for resolved galaxies. 
However, their method is not adopted in all of the TRGB observations
collected in Table 2: these results come from different authors, and have been obtained
by adopting different statistical procedures for determining the TRGB position.

Keeping in mind these warnings about the problems and the 
heterogeneity of the $I_{TRGB}$ observations, it is a safe conclusion to say
that in the limit of the accuracy of the present determinations of $I_{TRGB}$
in resolved galaxies, the comparison between theoretical TRGB distance 
scale and Cepheid distances with the zero point set by a LMC distance modulus
equal to 18.50 mag shows a statistically significant systematic difference, TRGB
distances being higher.

Since the existence of this difference, we also performed a test for assessing whether
it could be correlated with the metallicity of the Cepheids. 
Since the Madore \& Freedman (1991)
calibration of the P-L relation is independent of [M/H],
the presence of such a correlation could be
an indication of the need to correct the Cepheid distances also 
for metallicity effects (but see also the discussion in De Santis 1997 and 
Madore \& Freedman 1998). For example, in a recent analysis of the P-L
relation of a sample of 481 Cepheids in the LMC and SMC, 
Sasselov et al. (1997) derived a linear relation between
the distance modulus correction ($\delta\mu$) 
to apply to the distances computed with a metallicity-independent 
calibration, and the metallicity of the Cepheid population, of the form
$\delta\mu$=0.4([Fe/H]+0.3).

The problem in performing this comparison is the lack of direct
determinations of Cepheids metallicities for the sample of galaxies
in Table 2, with the only exception of the LMC.
Of course, it is not correct to use the metallicities derived from the RGB 
stars (reported in column 4 of Table 2), since 
they are typical of old stellar populations, while the Cepheids are
much younger stars.

To solve this problem we have decided to adopt, as representative
of the Cepheids original chemical composition, the [O/H] determinations
for the HII regions of the parent galaxies. 
We have restricted our analysis to the 7 dwarf irregular galaxies
(LMC, NGC3109, NGC6822, IC1613, WLM, Sex A and Sex B)
within the sample displayed in Table 2; this because the presence of
metallicity gradients in the disk of spiral galaxies, while
(as discussed in Skillman et al.1989) in dwarf irregular galaxies
it seems that the dispersion in the [O/H] abundances of HII regions
is quite small.
Moreover, as discussed by Luck \& Lambert (1992) in the case of
the LMC and SMC, 
the [O/H] values determined from the HII regions agree well with
the [O/H] values determined for the Cepheids and supergiants.
In the case of the LMC they provide for the Cepheids and supergiants
[O/H]=-0.70, in good agreement with the value
[O/H]=-0.58 as derived from the LMC HII regions by Skillman et al. (1989).

Figure 9 displays the difference $(m-M)_{TRGB}-(m-M)_{Cepheid}$
(obtained from Table 2 after averaging the values corresponding to
the three different sets of $BC_{I}$)
as a function of [O/H] for the Cepheids in the 7 irregular galaxies
previously quoted, spanning a range of almost 1 dex in [O/H].
The empty circle indicates the point corresponding to Sex B if a
Cepheid distance modulus of 25.82 mag is assumed (see previous
discussion); the typical error on [O/H] is of $\approx$0.20 dex.

The (O/H) ratios for the 7 irregular galaxies are from 
Skillman et al. (1989), with the exception of NGC3109 for which we used
the results by Hunter \& Gallagher (1985). The [O/H] values are
derived assuming (as in Skillman et al. 1989) 12+log$(O/H)_{\odot}$=8.92 
(Lambert 1978). Of course, when using the quantity [O/H] 
for obtaining the correct (qualitatively and quantitatively) 
ranking in metal content for the Cepheids in the galaxies in our sample, we are
assuming  that the [O/Fe] ratio is the same 
in all the galaxies considered, irrespective of the 
[O/H] absolute value.

We have performed a simple statistical analysis with the data
displayed in Figure 9, computing the linear correlation coefficient
$r_{a}$ between $(m-M)_{TRGB}-(m-M)_{Cepheid}$ and [O/H]. 
To be conservative, 
we have chosen to accept the existence of a linear relation 
between these two quantities only when the probability $P$ to derive
a value r$\geq r_{a}$ from a random sample 
of ($(m-M)_{TRGB}-(m-M)_{Cepheid}$,[O/H]) values
is less than 5\%.
Adopting for Sex B the Cepheid distance modulus displayed in Table 2
we find $P\approx$15\%, while $P\approx$5\% if a Cepheid
distance modulus of 25.82 mag is adopted.
We have also performed the same test excluding Sex B from the sample;
in this case we have considered only 6 galaxies, obtaining again
$P\approx$15\%.

Due to the dependence of the result of this simple statistical
analysis on the distance to Sex B, and to the quite large error bars
associated to the observational points, we conclude that,
in the limit of the small sample considered and of our assumptions on
the Cepheids metal content, there is no clear evidence for a
a linear correlation between $(m-M)_{TRGB}-(m-M)_{Cepheid}$ and
Cepheid metallicity.

\section{\bf The Leo I group TRGB distance and the value of $H_{0}$}

\tx

Using the TRGB distance scale set by our evolutionary models,
we now determine the value of $H_{0}$ in the same way as recently 
done by Tanvir et al (1995), SA97, Thomsen et al (1997), 
Hjorth \& Tanvir (1997), Gregg (1997).
This means that at first we determine the TRGB distance 
to the Leo I group of galaxies. This group is much more compact than,
for instance, the Virgo cluster and from this point of view
it is better suited for a determination of this important cosmological
parameter. However, also if the Leo I group is not affected by the
uncertainty in distance introduced by the depth of the Virgo cluster,
it is not sufficiently far away for the local peculiar velocity field
being only a small fraction of its recession velocity.
Therefore we will use the Leo I distance as the zero point for the
distance to the Coma cluster, that is sufficiently far away for
determining $H_{0}$ with only a small
indetermination due to the local velocity field.
The distance to the Coma cluster will be determined adopting 
purely differential estimates (by means of secondary distance
indicators) of the relative distance Leo I-Coma.

\subsection{\bf The TRGB distance to Leo I}

\tx

The Leo I group consists of five dominant galaxies: NGC3351 [M95,SB(r)b]
NGC3368 [M96, SAB(rs)ab], NGC3377[E5/6], NGC3379 [E1], NGC3384 [SB(s)0]; it
is relatively nearby, compact, with a full line-of-sight depth estimated to
be $\approx$8\% compared to its distance, assuming spherical symmetry
(Tanvir et al.1995). Its mean recession velocity is of the order of 700 Km/s (SA97).
Schneider (1989) shows a strong argument in favour of the five
dominant galaxies being in close physical proximity, probably all within 0.5 Mpc or less
from each other. In particular the group contains a ring of
intergalactic neutral hydrogen which is orbiting around the close pair
NGC3379 and NGC3384, and seems to be interacting with NGC3368.

Very recently, SA97 detected the TRGB in NGC3379, by
means of $HST$ WFPC2 observations. They placed the observed TRGB at
I=26.32$\pm$0.05 mag, assumed $A_{I}$=0.02 mag (Burstein \& Heiles 1982), a negligible
internal reddening in NGC3379 (due to the location $6^{'}$ west from the
NGC3379 nucleus of the target field), and a metallicity
[M/H]=-0.68$\pm$0.40 as derived 
from the $(B-I)$ color at the target field according to Sodeman \& Thomsen
(1994), using the relation between [M/H] and $\rm (B-I)$ by Couture et al. (1990).

By adopting the quoted  values for extinction, metallicity and TRGB
location, we derive a distance modulus  $(m-M)_{0,3379}$=30.44$\pm$0.13 mag 
when using the DA90 $BC_{I}$,
$(m-M)_{0,3379}$=30.53$\pm$0.09 mag when using Equation 6 (Yale transformations)
and $(m-M)_{0,3379}$=30.48$\pm$0.13 mag when using Equation 5 (K97 transformations).
It is worth noticing that the distance modulus obtained by adopting the DA90
$BC_{I}$ is 0.14 mag higher than the distance modulus derived by SA97.
This difference is due exclusively to the use of the updated TRGB brightness-[M/H]
relation given by Equation 1.

The error budget used in
deriving the uncertainty on the distance to NGC3379 is reported in Table 4. 
The contributions corresponding to the
different error sources are added in quadrature to obtain the total
uncertainty. 

At this point, by considering the difference among the three 
NGC3379 distance moduli as an indication of the error due to the
uncertainty associated to the bolometric correction scale, we adopt a final value
$(m-M)_{0,3379}$=30.46$\pm$0.16 mag, that spans all the range of
distance moduli allowed by the three sets of $BC_{I}$.

This distance modulus corresponds to a linear distance
$d_{3379}$=12.4$\pm$0.9 Mpc, that is $\approx$8\% higher than the
value derived by SA97. 

\table{4}{S}{\bf Table 4. \rm Error budget used in deriving the uncertainty on the 
distance to NGC3379. The three different values for the error due to 
the uncertainty on the metallicity correspond to the three different
sets of bolometric corrections used. The errors on the galactic extinction, TRGB measurement
and WFPC2 photometric zero point come from SA97.}
{\halign{%
\rm#\hfil&\hskip10pt\hfil\rm#\hfil&\hskip10pt\hfil\rm\hfil#&\hskip10pt\hfil\rm\hfil#\cr
Source & Error(mag)& &   \cr 
\noalign{\vskip 10pt}
Galactic extinction     &$\pm$0.02       &           &          \cr 
TRGB measurement        &$\pm$0.05       &           &          \cr
Photometric zero point  &$\pm$0.04       &           &          \cr   
[M/H] ($\pm$0.40 dex)   &$\pm$0.09(DA90)&$\pm$0.03(Yale)&$\pm$0.09(K97)\cr
Y  ($\pm$0.03)          &$\pm$0.03       &           &          \cr
$BC_{I}$ zero point     &$\pm$0.02       &           &          \cr 
Theoretical calibration &$\pm$0.05       &           &          \cr
         &                &           &          \cr
Total uncertainty       &$\pm$0.13(DA90)&$\pm$0.09(Yale)&$\pm$0.13(K97)\cr}}       

It is now interesting to compare our derived distance to the Leo I group 
with the results from Cepheid observations. 
By adopting the HST observations of 7 Cepheid stars in NGC3368 by Tanvir et al. (1995),
using the Madore \& Freedman (1991) calibration (the same used in
the previous section), and by correcting for the photometric zero
point as discussed in SA97, the Cepheid distance
modulus to Leo I is $(m-M)_{0,3368}$=30.36$\pm$0.13 mag 
(the error budget is derived from Tanvir et al. (1995) by excluding the
contribution due to the possible systematic error on the LMC distance,
since this is exactly what we are trying to determine in the present study). 
The difference between the best value of the distance to Leo I as
derived from the Cepheids and our analogous quantity as derived
from the TRGB, is in agreement with the result of the same comparison 
performed with the sample of 12 galaxies listed in Table 2.

More recently Graham et al. (1997) discovered, through {\sl HST}
observations, 49 probable Cepheids in NGC3351; their derived distance
modulus (adopting the  Madore \& Freedman (1991) calibration)
is 30.01$\pm$0.16 mag (once again, the error does not include the
contribution due to the uncertainty on the distance of the LMC).
This value is much lower than distance derived from the TRGB, and also
lower than the NGC3368 Cepheid distance.
This difference in the Cepheid distance 
moduli of NGC3351 and NGC3368 would imply a reciprocal distance of $\approx$
1.8 Mpc, difficult to reconcile, as already discussed by Gregg (1997),
with the results by Schneider (1989).
Waiting for other results about Cepheids in the Leo I group, we consider 
this discrepancy as a measure of the uncertainty on the present
distance estimates by means of the Cepheid P-L relation.

\subsection{\bf From Leo I to Coma and the value of $H_{0}$}

\tx

The second step toward the determination of $H_{0}$ is the relative
distance between Leo I and Coma.
According to the recent analysis by Colless \& Dunn (1996), the Coma
cluster is likely to consist of two components: the main cluster 
centered around NGC4874 and NGC4889, with a mean 
recession velocity $cz$=6853 km $s^{-1}$ and a subgroup around NGC4839
characterized by a mean value of $cz$=7339 km $s^{-1}$. 
The virial mass of the main cluster results to be around one order of
magnitude higher than the subgroup mass.

By adopting a relative distance between the Coma cluster main component and Leo I 
$(m-M)_{0,Coma}$-$(m-M)_{0,Leo I}$=4.73$\pm$0.13 mag as derived from 
the diameter-velocity dispersion data by Faber et al. (1989) and
$cz$=6853$\pm$100 km $s^{-1}$ ,as done by SA97, we obtain:
$$(m-M)_{0,Coma}=35.19\pm0.21 \rm mag $$
\noindent
and a linear distance  $d_{Coma}$=109$\pm$10 Mpc (in the error budget it has been 
taken into account the error on the Leo I distance modulus, the
uncertainty on the relative distance Coma-Leo I and
an error of $\pm$0.04 mag that takes into account the r.m.s. depth 
of the Leo I group as adopted by Tanvir et al. (1995)), and finally
$H_{0}$=63$\pm$6 Km $s^{-1} Mpc^{-1}$.
This value is lower by 
5 Km $s^{-1} Mpc^{-1}$ than the value derived by SA97
with the same method, the same observational data for Leo I, the same
relative distance Coma-Leo I, the same recession velocity for the Coma
cluster, but an old calibration of the TRGB theoretical luminosities.

However, the recession velocity for the Coma cluster, adopted by 
SA97, is the value given by Colless \& Dunn (1996), which corresponds to the heliocentric
recession velocity, not to the cosmologic recession velocity.
We have therefore transformed the heliocentric recession velocity to the
centroid of the Local Group and corrected for the motion of the Local
Group relative to the cosmic background radiation in the direction of Coma
(272 Km $s^{-1}$ according to Staveley-Smith \& Davies (1989), to which we
attribute an error by $\pm$100 Km $s^{-1}$). Moreover, we have corrected for
the peculiar motion ($V_{p}$) of the cluster as estimated by Han \&
Mould (1992): $V_{p}$=+66$\pm$428 Km $s^{-1}$ (the median value of their three solutions for 
$V_{p}$ has been adopted). So a cosmic recession velocity for the Coma 
cluster equal to $cz$=7068$\pm$440 km $s^{-1}$ is finally obtained.

Moreover, we have searched in the literature for recent independent
determinations of the relative distance Coma-Leo I: the most recent results
are by Gregg (1997), Thomsen et al (1997) and Hjorth \& Tanvir (1997).

Gregg (1997) determined diameter-velocity dispersion relations in B,V,K bands 
for NGC3377, NGC3379 and NGC3384 in the Leo I group from published photometry and kinematic
data. These relations, whose slopes in the three colors are in good agreement with
those for 24 galaxies in the 
main component of the Coma cluster, yield an estimate of the Coma-Leo I distance ratio 
$d_{Coma}/d_{Leo I}$=8.84$\pm$0.23. This value, coupled with our TRGB
distance to the Leo I group gives a linear distance $d_{Coma}$=110$\pm$11 Mpc.
Thomsen et al. (1997)
applied the surface brightness fluctuations technique for
deriving the relative distance between NGC3379 and NGC4881 in the main
component of the Coma cluster. 
They derived $(m-M)_{0,Coma}$-$(m-M)_{0,Leo I}$=4.89$\pm$0.30 mag,
which gives $(m-M)_{0,Coma}$=35.35$\pm$0.34 mag and a linear distance
$d_{Coma}$=117$\pm$20 Mpc.
Hjorth \& Tanvir (1997) determined a distance ratio
$d_{Coma}/d_{Leo I}$=9.5$\pm$0.7 through the construction of the
fundamental plane of the Leo I group. This distance ratio provides 
$d_{Coma}$=118$\pm$18 Mpc.

Considering the cosmic recession velocity previously given:
$cz$=7068$\pm$440 km $s^{-1}$, these three distances give
values of the Hubble constant equal to $H_{0}$=64$\pm$7, 60$\pm$11 Km $s^{-1} Mpc^{-1}$
and 60$\pm$10 Km $s^{-1} Mpc^{-1}$, respectively, in good agreement within each other.

\section{\bf Summary and conclusions}

\tx

The main results presented in this paper can be summarized as follows:
\medskip\noindent
1) we have presented theoretical relations between $M^{tip}_{Bol}$
or $M_{I}^{tip}$ and metallicity, covering the range 
$-2.35\le[M/H]\le-0.28$. These relations, obtained from evolutionary
stellar models computed with updated input physics, expand the metallicity range
covered by the relations presented in Paper I and by the old
calibration by LFM93. We use the empirical $BC_I$ values from DA90,
the semiempirical $BC_I$ from the Yale transformations and the purely
theoretical ones from K97 when deriving galaxies
distance moduli from their TRGB I (Cousins) magnitude.
A particular attention has been devoted to the correct 
calibration of the zero point of the bolometric correction
scales.
\medskip\noindent
2) the ZAHB models presented in Cassisi \& Salaris (1997) have been transformed to
the observational V-$\rm (B-V)$ plane by adopting the Yale and K97
transformations. The relations obtained adopting these two
sets of transformations agree quite well;
\medskip\noindent
3) the GC distance scale set by the HIPPARCOS subdwarfs agrees well,
within the errors, with the distance scale set by our ZAHB models for
RR Lyrae stars. At the same time TRGB and ZAHB distance scales for GC
agree within the statistical uncertainties associated to the location of the
TRGB in GC; 
\medskip\noindent
4) with the distance scale set by our ZAHB models, we have
recalibrated the relation [M/H]-$(V-I)_{0,-3.5}$ given in Paper I;
\medskip\noindent
5) when considering stellar populations in resolved galaxies,
the RR Lyrae distance scale agrees well with the TRGB; the 
mean difference between the two distance scales is statistically compatible
with a value equal to zero;
\medskip\noindent
6) when using the available observational sample of 
TRGB determinations in resolved galaxies, 
the comparison between TRGB and  
Cepheid  distance scales suggests the revision of the
Madore \& Freedman (1991) zero point for the Cepheid distances, that is
set by a LMC distance modulus equal to 18.50 mag. In particular, 
we find that TRGB distances
are on average 0.12 mag larger than the ones determined from the Cepheids.
We could not find any compelling evidence that this difference is linearly
correlated with the metallicity of the Cepheid population;
\medskip\noindent
7) we have determined the distance to NGC3379 in the Leo I group by
using the observational TRGB data by SA97 and our
theoretical TRGB models. We obtain a distance of 12.4$\pm$0.9 Mpc,
that is 8\% higher than the value obtained by SA97
using the old LFM93 calibration of the theoretical TRGB luminosities;
\medskip\noindent
8) we have used the most recent determinations of the relative distance
between Leo I and Coma cluster from Gregg (1997), Thomsen et al. (1997) and 
Hjorth \& Tanvir (1997), obtaining 
$d_{Coma}$=110$\pm$11, 117$\pm$20 and 118$\pm$18 Mpc, respectively;
\medskip\noindent
9) from these Coma cluster distances by assuming a cosmologic recession velocity
for the main component of the cluster $cz$=7068$\pm$440 km $s^{-1}$,
we obtain: $H_{0}$=64$\pm$7, 60$\pm$11 and 60$\pm$10 Km $s^{-1} Mpc^{-1}$, respectively.
The final error on $H_{0}$ is computed by taking into account many
different sources of errors in the evaluation of the Leo I and Coma distance moduli, 
and the uncertainty on the Coma recession velocity.
\smallskip
\noindent
These results indicate clearly that the use of the TRGB as a primary
distance indicator, calibrated by means of updated stellar models,
provides a distance scale systematically longer in comparison with 
the determinations obtained by adopting the older TRGB calibration by
LFM93 or the Cepheid distance scale by Madore \& Freedman (1991).
Moreover, the agreement between TRGB, ZAHB and HIPPARCOS subdwarfs distances
confortably assesses the reliability of our theoretical models.

What are the implications of these $H_{0}$ values for the age of
the universe?

To answer to this question we have to assume a value for the mass
density parameter $\Omega=\Omega_{m}+\Omega_{\Lambda}$, where the two
terms take into account respectively the contribution of the mass
density and of the vacuum energy density.
Independently of the constraints from the simple standard flat
inflationary models with $\Omega=\Omega_{m}$=1, 
the values of $\Omega_{m}$ as observationally determined adopting different methods 
suggest 0.3$\leq \Omega_{m} \leq$1.0 (see, e.g., Bartelmann et al. 1998,
Dekel 1997, Steigman et al. 1997, Perlmutter et al. 1997).
Assuming conservatively $\Omega_{\Lambda}$=0, 
and considering $H_{0}$=60$\pm$11 that reproduces
all the range of values for the Hubble constant previously given,  
we can consider the two extreme cases:
\smallskip\noindent
i) flat universe with $\Omega_{m}$=1;
$H_{0}$= 60$\pm$11 Km $s^{-1}$ $Mpc^{-1}$
provides an age of the universe t$\approx$11.0$\pm$2.5 Gyr;
\smallskip\noindent
ii) open universe with $\Omega_{m}$=0.3;
$H_{0}$= 60$\pm$11 Km $s^{-1} Mpc^{-1}$
gives t$\approx$13.0$\pm$2.5 Gyr.
\noindent
The use of a cosmological constant different from zero increases
in both cases the age of the universe with respect to the quoted values.

These results nicely agree with the most recent determinations of the
GC ages, also in the case of the
most restrictive constraint $\Omega_{m}$=1, $\Omega_{\Lambda}$=0. 
For instance, Salaris et al. (1997) and Salaris \& Weiss (1997)
find for the older GC $t_{GC}$$\approx$12$\pm$1 Gyr, while 
Gratton et al. (1997) and Chaboyer et al. (1997) obtain
respectively $t_{GC}$$\approx$11.8$\pm$2 Gyr and 
$t_{GC}$$\approx$11.5$\pm$1.3 Gyr.

This concordance between age of the GC and age of the universe
as derived by adopting a spectrum of reasonable choices of $\Omega$,
suggests that the long-standing conflict between the Hubble age and GC
ages is resolved when adopting updated stellar models for deriving the
GC ages and for calibrating the TRGB distance scale.

\section*{Acknowledgments}

\tx 

It is a pleasure to thank S.Schindler for illuminating discussions
about the properties of clusters of galaxies, M. Groenewegem for 
supplying his results about the LMC distance modulus in advance
of publication, and for useful suggestions about the
distance scale problem, Matthias Bartelmann and Doris Neumann for
discussions and useful references about recent determinations of $\Omega$.
A. Weiss is warmly thanked for encouragement and fruitful discussions during all 
the development of this work.
We warmly thank G. Bono, for reading a preliminary version of this paper and for
his helpful suggestions.
We are also grateful to F. Castelli for many interesting discussions and 
for providing updated bolometric corrections and color transformations.

\noindent
The referee, N. Tanvir, is warmly thanked for many useful and pertinent 
remarks that improved the quality of the paper.

\noindent
The work of one of us (M.S.) was carried out as part of the TMR Programme
(Marie Curie Research Training Grants) financed by the EC.

\section*{References}
\bibitem Alonso A., Arribas S. \& Martinez-Roger C 1995, A\&A, 297, 197
\bibitem Bartelmann M., Huss A., Colberg J.M., Jenkins A. \& Pearce
F.R. 1998, A\&A, 330, 1
\bibitem Burstein D. \& Heiles C. 1982, AJ, 87, 1165
\bibitem Burstein D. \& Heiles C. 1984, ApJS, 54, 33
\bibitem Caloi V., D'Antona F. \& Mazzitelli I. 1997, A\&A, 320, 823
\bibitem Cardelli J.A., Clayton G.C. \& Mathis J.S. 1989, ApJ, 345, 245
\bibitem Carigi L., Colin P., Peimbert M. \& Sarmiento A. 1995, ApJ, 445, 98
\bibitem Cassisi S. \& Salaris M. 1997, MNRAS, 285, 593
\bibitem Cassisi S., Castellani V., Degl'Innocenti S. \& Weiss
A. 1997, A\&A, {\sl in press}
\bibitem Castelli F \& Kurucz R.L. 1994, A\&A, 281, 817
\bibitem Castelli F., Gratton R.G. \& Kurucz R.L. 1997a, A\&A, 318, 841
\bibitem Castelli F., Gratton R.G. \& Kurucz R.L. 1997b, A\&A, 324, 432
\bibitem Chaboyer B., Demarque P., Kernan P.J. \& Krauss L.M. 1997,
ApJ, {\sl submitted to}
\bibitem Code A.D., Davis J., Bless R.C. \& Hanbury Brown R. 1976, ApJ, 203, 417
\bibitem Colless M. \&  Dunn A.M. 1996, ApJ, 458, 435
\bibitem Couture  J., Harris W.E. \& Allright J.W.B. 1990, ApJ, 73, 671
\bibitem Da Costa G.S. \& Armandroff T.E. 1990, AJ, 100, 162
\bibitem Dekel A. 1997, in 'Galaxy Scaling Relations: Origins,
Evolution and Applications', ed. L. Da Costa (Springer) {\sl in press}
\bibitem De Santis R. 1996, A\&A, 306, 755
\bibitem De Santis R. 1997, A\&A, {\sl submitted to}
\bibitem Elson R. A. W. 1996, MNRAS, 286, 732
\bibitem Faber S.M. et al.1989, ApJS, 69, 763
\bibitem Feast M.W. \& Catchpole R.M. 1997, MNRAS, 286, L1
\bibitem Frogel J.A., Persson S.E. \& Cohen J.G. 1983, ApJS, 53, 713
\bibitem Graham J.A. et al.1997, ApJ, 477, 535
\bibitem Gratton R.G., Fusi Pecci F., Carretta E., Clementini G.,
Corsi C.E. \& Lattanzi M. 1997, ApJ, 491, 749
\bibitem Green E.M. 1988, in "Calibration of Stellar Ages", A.G. Davis Philip ed. 
(L. Davis Press) p. 81
\bibitem Gregg M.D. 1997, New Astronomy, 1, 363
\bibitem Han M. \& Mould J. 1992, ApJ 396, 453
\bibitem Hayes D.S., IAU Symp. 111, p.225
\bibitem Hunter D.A. \& Gallagher J.S. 1985, ApJS, 58, 533
\bibitem Hjorth J. \& Tanvir N.R. 1997, ApJ, 482, 68
\bibitem Kent S.M. 1985, PASP, 97, 165
\bibitem Kurucz R.L. 1979, ApJS, 40, 1
\bibitem Lambert D.L. 1978, MNRAS, 182, 249
\bibitem Lee M.G., Freedman W. \& Madore B.F. 1993, ApJ, 417, 553
\bibitem Lee Y.-W., Demarque P. \& Zinn R. 1990, ApJ, 350, 155
\bibitem Luck R.E. \& Lambert D.L. 1992, ApJS, 79, 303
\bibitem Madore B. F. \& Freedman W.L. 1991, PASP, 103, 933
\bibitem Madore B. F. \& Freedman W.L. 1995, AJ, 109, 1645
\bibitem Madore B. F. \& Freedman W.L. 1998, ApJ, 492, 110
\bibitem Musella I., Piotto G. \& Capaccioli M. 1997, AJ, 114, 976
\bibitem Oudumaijer R.D., Groenewegen M.A.T. \& Schrijver H. 1997,
MNRAS, {\sl in press}
\bibitem Peimbert M. 1993, Rev. Mex. Astr. Astrofis., 27, 9
\bibitem Perlmutter S. et al.1997, ApJ, 483, 565
\bibitem Reid I.N. 1997, AJ, 114, 161
\bibitem Sakai S., Madore B.F. \& Freedman W.L. 1996, ApJ, 461, 713
\bibitem Sakai S., Madore B.F., Freedman W.L., Lauer T.R., Ajhar E.A.,
Baum W.A. 1997a, ApJ, 478, 49
\bibitem Sakai S., Madore B.F. \& Freedman W.L. 1997b, ApJ, 480, 589
\bibitem Salaris M. \& Cassisi S. 1996, A\&A, 305, 858
\bibitem Salaris M. \& Cassisi S. 1997, MNRAS, 289, 406
\bibitem Salaris M. \& Weiss A. 1997, A\&A 327, 107
\bibitem Salaris M. \& Weiss A. 1998, {\sl in preparation}
\bibitem Salaris M., Chieffi A. \& Straniero O. 1993, ApJ, 414, 580
\bibitem Sasselov D.D. et al. 1997, A\&A, 324, 471
\bibitem Schneider S.E. 1989, ApJ 343, 94
\bibitem Skillman E.D., Kennicutt R.C. \& Hodge P.W. 1989, ApJ, 347, 875 
\bibitem Sodeman M. \& Thomsen B. 1994, A\&A, 292, 425
\bibitem Soria R. et al. 1996, ApJ, 465, 79
\bibitem Staveley-Smith L. \& Davies R.D. 1989, MNRAS, 241, 787 
\bibitem Steigman G., Hata N. \& Felten J.E. 1997, {\sl preprint astro-ph/9708016}
\bibitem Straniero O., Chieffi A. \& Limongi M. 1997, ApJ, 490, 425
\bibitem Sweigart A. V. \& Gross P. G. 1978, ApJS, 36, 405
\bibitem Tanvir N.R., Shanks T., Ferguson H.C. \& Robinson D.R.T. 1995, Nature, 377, 27 
\bibitem Thomsen B., Baum W.A., Hammergren M. \& Worthey G. 1997, ApJ, 483, L37
\bibitem Vandenberg D.A. \& Bell R.A. 1985, ApJS, 58, 561
\bibitem Wheeler J.C., Sneden C. \& Truran J. W. 1989, ARAA, 252, 179
\bibitem Whitelock P.A., van Leeuwen F. \& Feast M.W. 1997, {\sl preprint astro-ph/9706096}
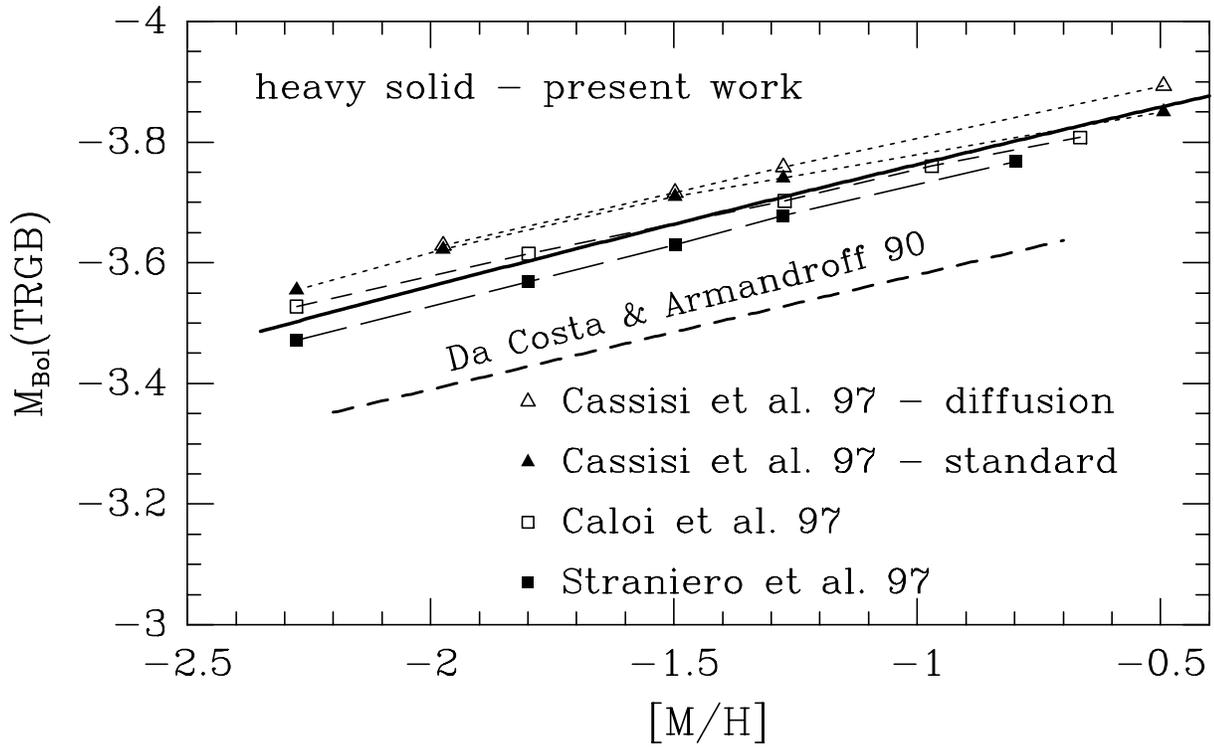
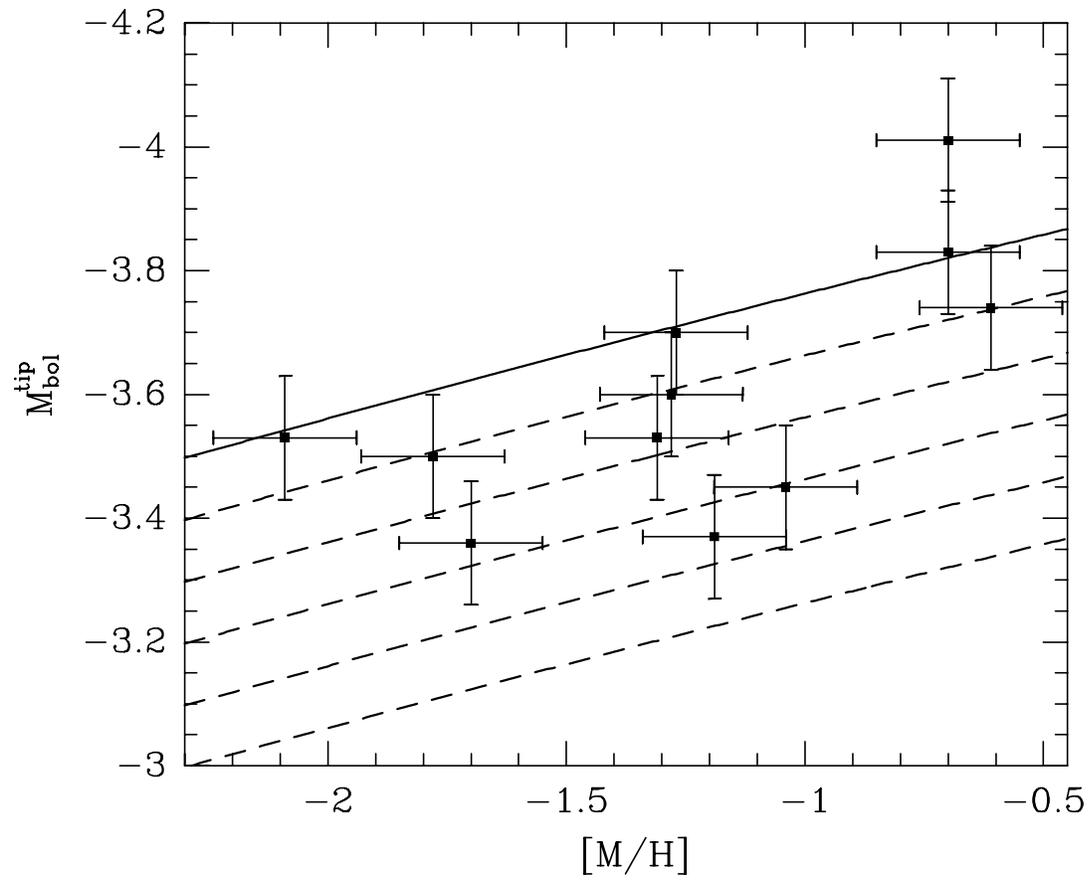
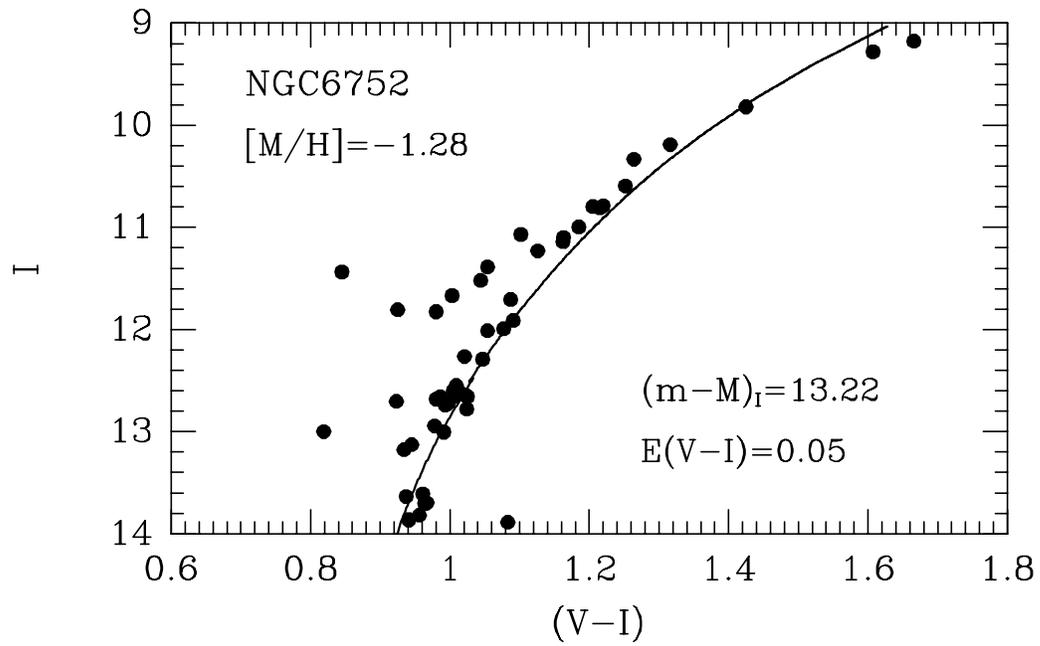
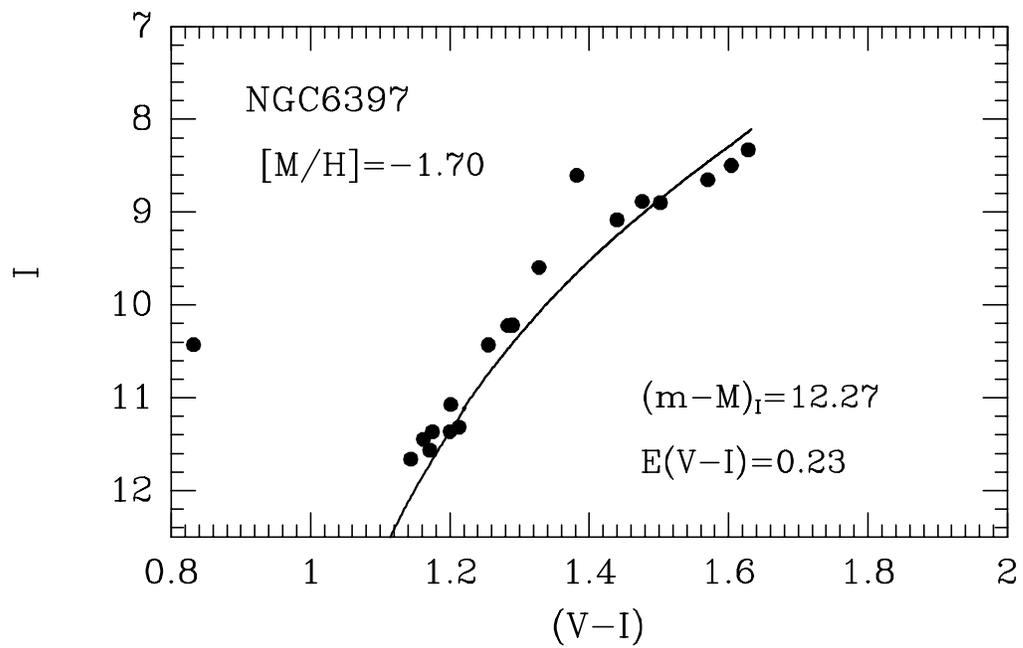
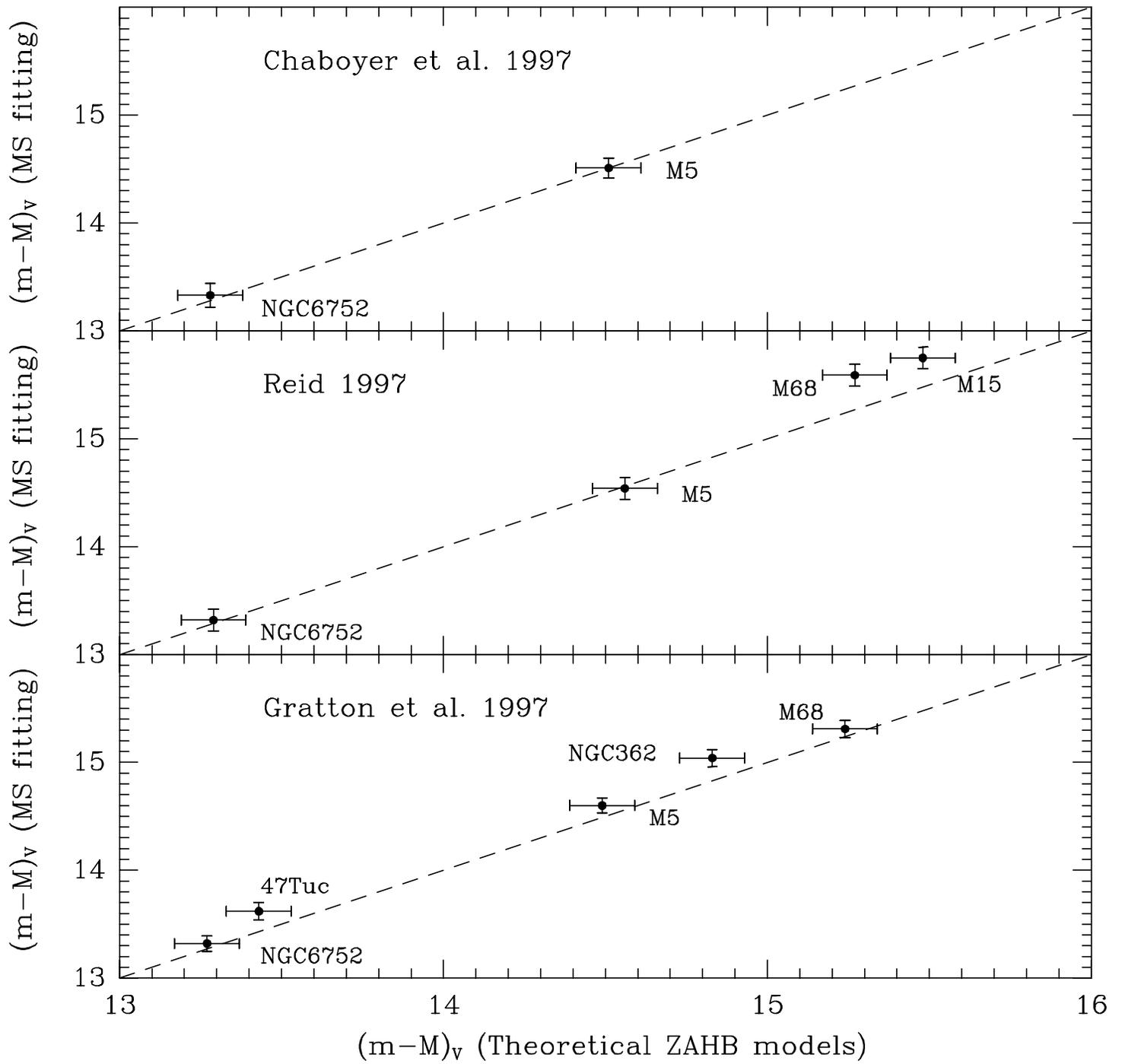
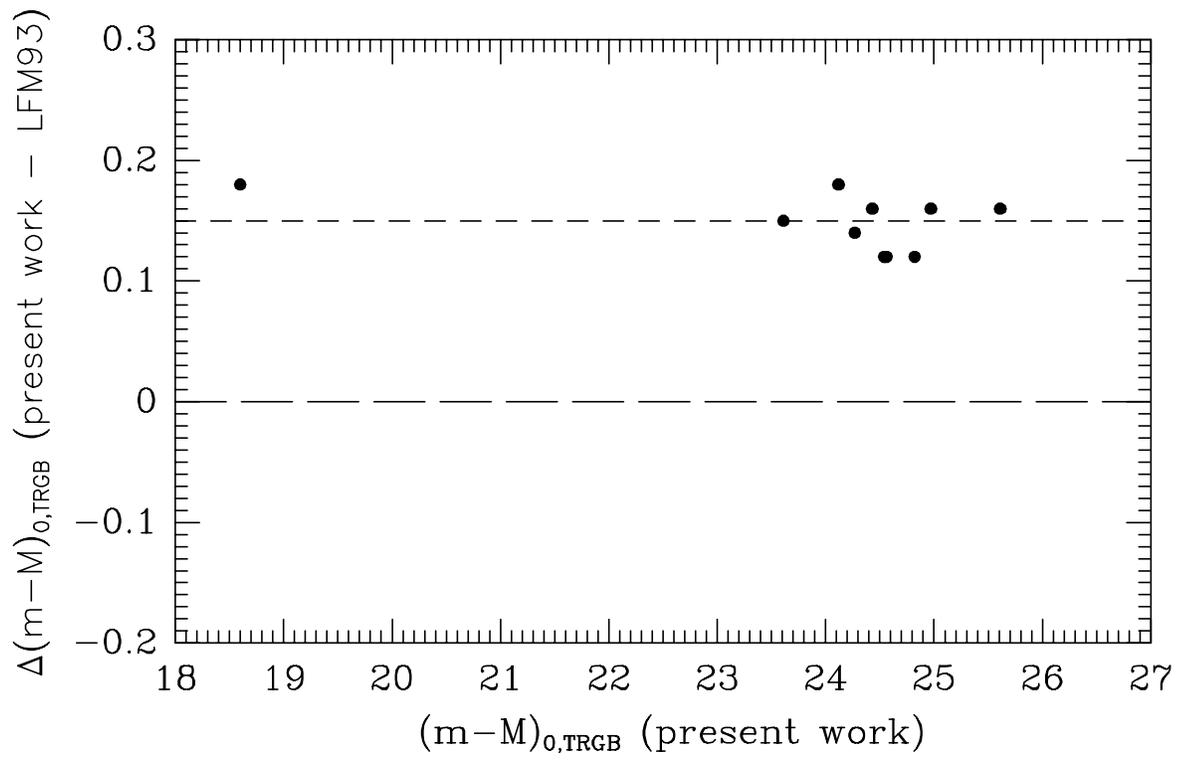
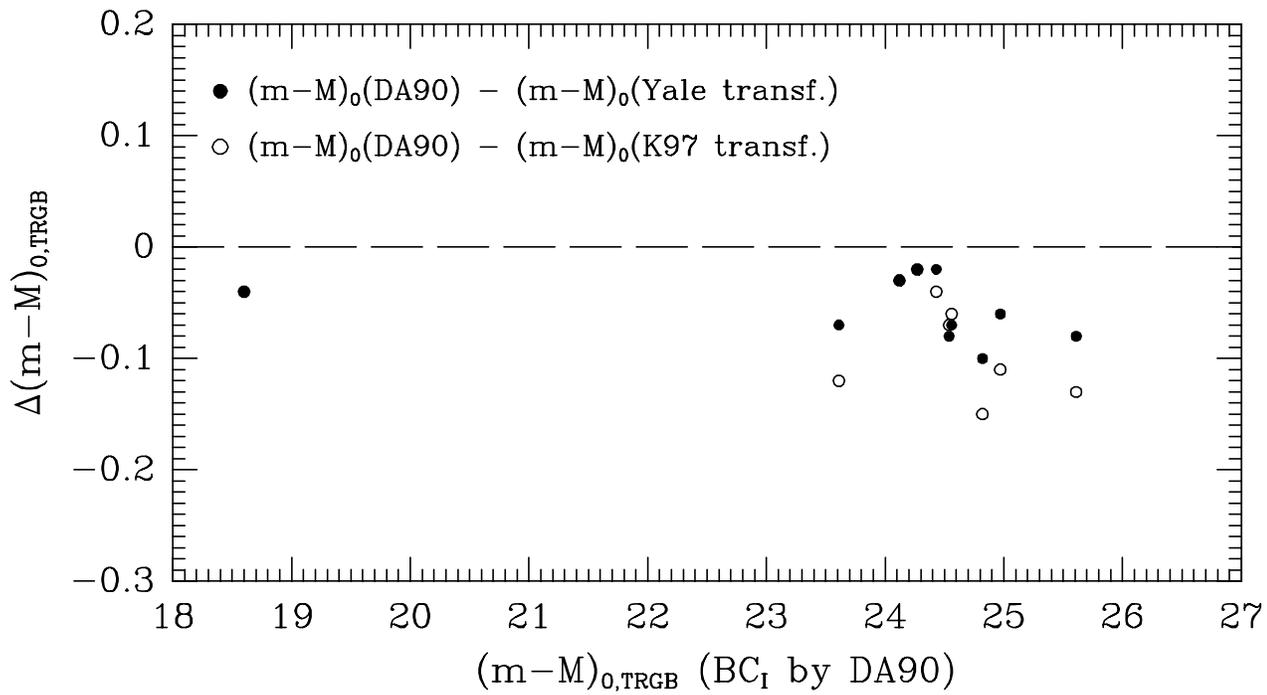
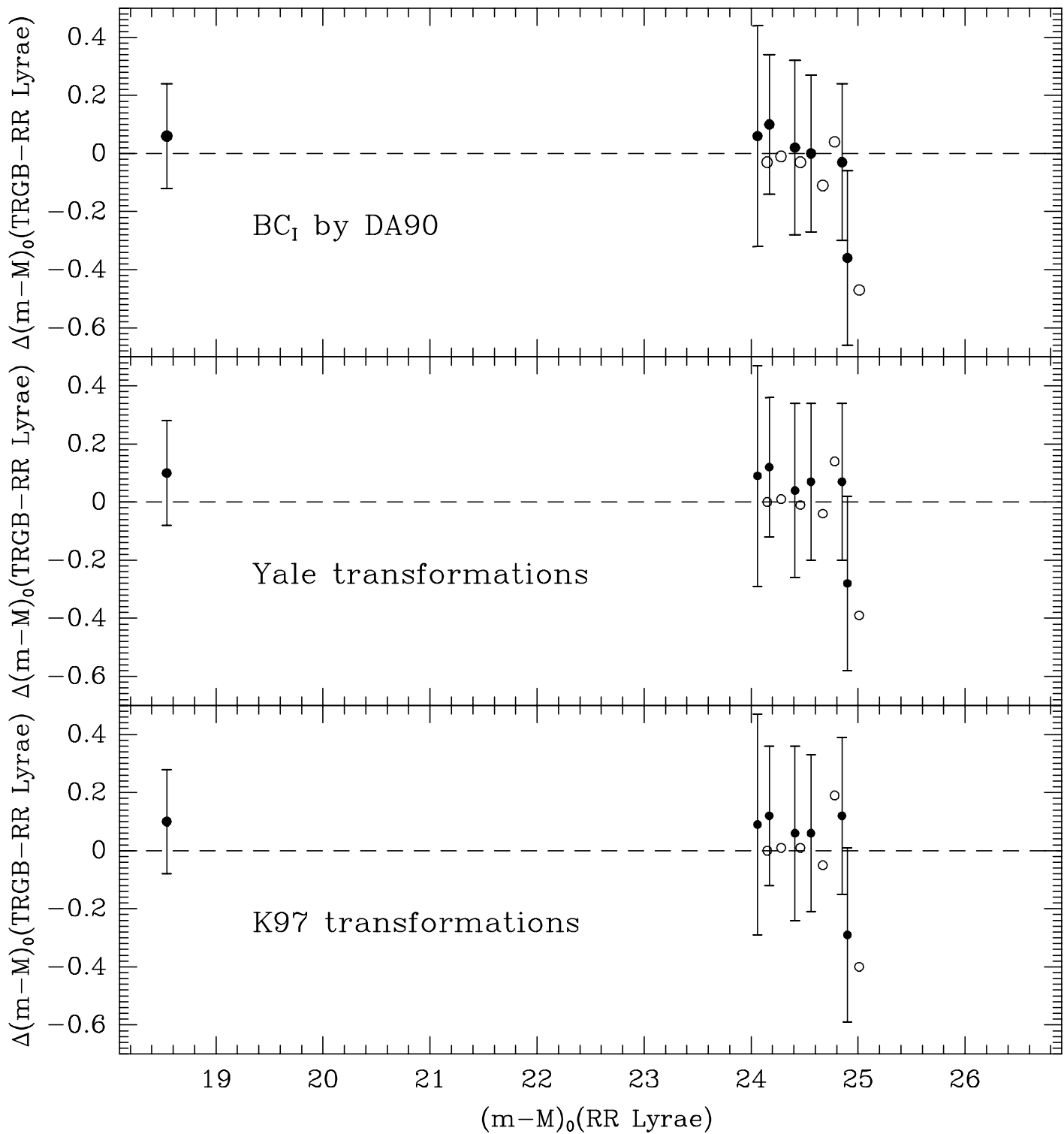
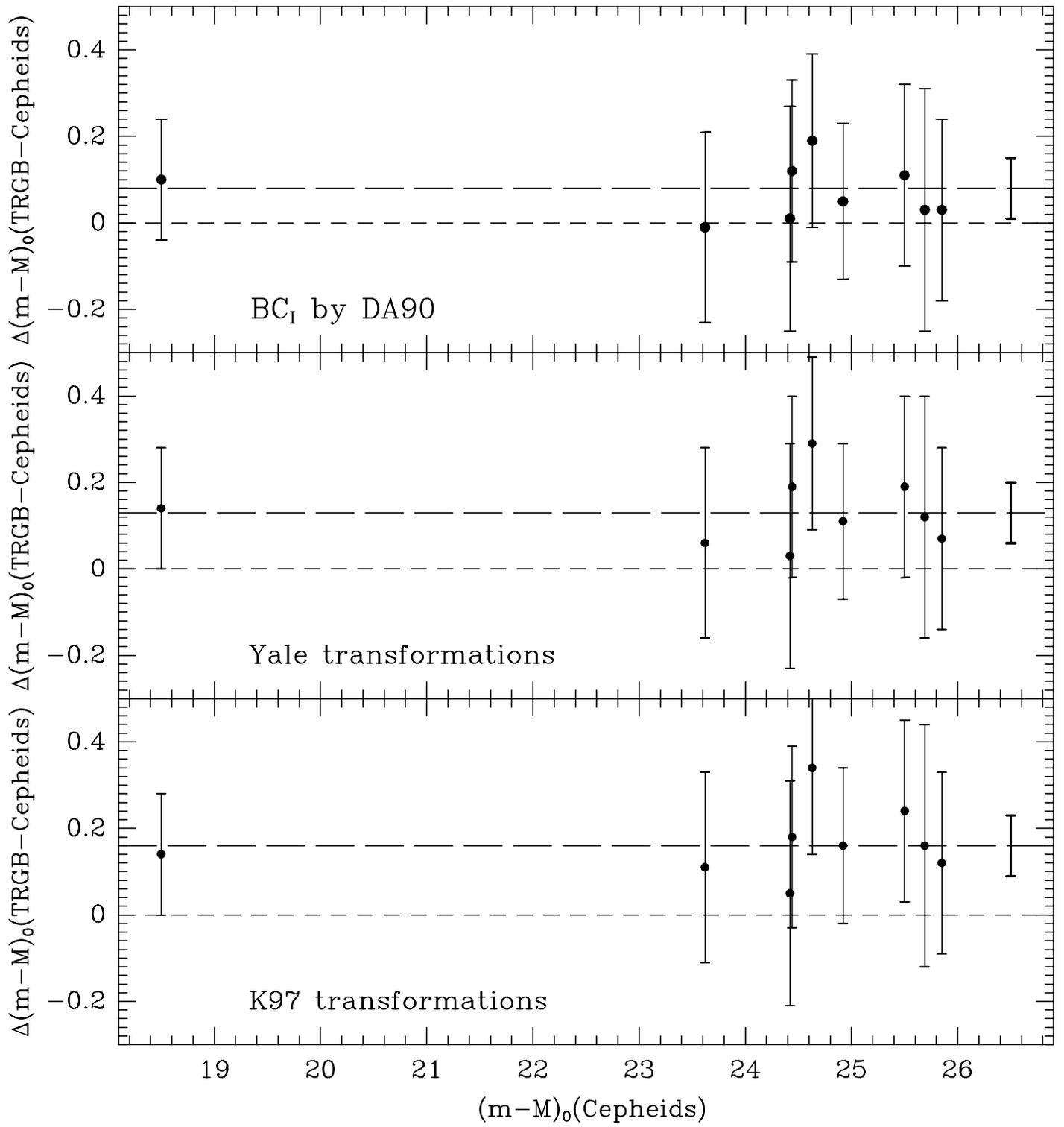
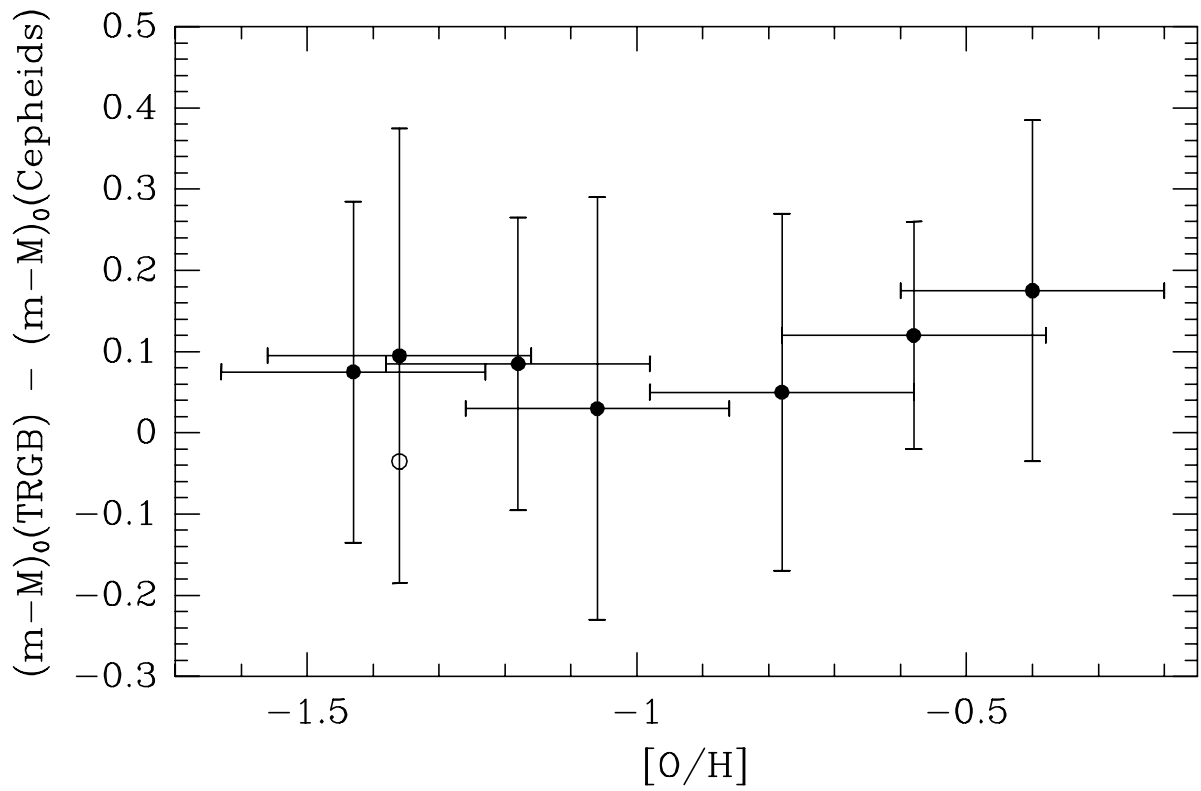
\bye